
\input phyzzx

\def\CMP#1{{\sl Comm. Math. Phys. {\bf #1}}}
\def\IMPA#1{{\sl Int. J. Mod. Phys. {\bf A#1}}}

\def\JSP#1{{\sl J.\ Stat. \ Phys.\ {\bf #1}}}
\def\JPA#1{{\sl J.\ Phys.\ {\bf A#1}}}

\def\LMP#1{{\sl Lett.\ Math.\ Phys.\ {\bf #1}}}
\def\MPL#1{{\sl Mod.\ Phys.\ Lett. \ {\bf #1}}}

\def\NPB#1{{\sl Nucl.\ Phys.\ {\bf B#1}}}

\def\PLB#1{{\sl Phys.\ Lett.\ {\bf #1B}}}

\def\TMP#1{{\sl Theor.\ Math.\ Phys.\ {\bf #1}}}


\def\nxl{\hfill\break}

\def\B{{\cal B}}
\def\D{{\cal D}}
\def\M{{\cal M}}                            
\def\U{{\cal U}}                            

\def\a{\alpha}

\def\b{\beta}
\def\g{\gamma}

\def\e{\epsilon}

\def\l{\lambda}

\def\s{\sigma}

\def\t{\theta}

\def\Th{\Theta}

\def\o{\over}

\def\bold#1{\setbox0=\hbox{$#1$}
     \kern-.025em\copy0\kern-\wd0
     \kern.05em\copy0\kern-\wd0
     \kern-.025em\raise.0433em\box0 }
\def\lowmp{\lower.11em\hbox{${\scriptstyle\mp}$}}

\def\Im{{\rm Im\,}}
\def\Re{{\rm Re\,}}

\def\sub#1#2{{{#1^{\vphantom{0000}}}_{#2}}}
\def\frac#1#2{{\textstyle{
 #1 \over #2 }}}                            


\def\ZZ{{\rm Z \!\! Z}}                       
\def\1{{\rm 1 \!\!\, l}}                        
\def\bra#1{\left\langle #1\right|}               
\def\ket#1{\left| #1\right\rangle}               
%

%
%


\hyphenation{Di-par-ti-men-to}
\hyphenation{na-me-ly}
\hyphenation{al-go-ri-thm}
\hyphenation{pre-ci-sion}
\hyphenation{cal-cu-la-ted}

\def\nxl{\hfill\break}
\def\B{{\hat {\cal B}}}
\def\D{{\hat {\cal D}}}
\def\Ns{{\scriptstyle N}}
\def\sub#1#2{{{#1^{\vphantom{0000}}}_{#2}}}
\Pubnum={$\rm PAR\; LPTHE\; 93/17
         \qquad {\rm march  \; 1993}$}
\date={}
\titlepage
\title{BETHE ANSATZ AND QUANTUM GROUPS}
\author{ H.J. de Vega }
\address{ Laboratoire de Physique Th\'eorique et Hautes Energies
     \foot{Laboratoire Associ\'e au CNRS UA 280}, Paris
     \foot{mail address: \nxl
           L.P.T.H.E., Tour 16 $1^{\rm er}$ \'etage, Universit\'e Paris VI,\nxl
           4 Place Jussieu, 75252, Paris cedex 05, France }}
\author{Based on Lectures given at the Vth. Nankai Workshop, Tianjin, P. R.
of CHINA, June  1992}
\endpage
\title{BETHE ANSATZ AND QUANTUM GROUPS}
\author{ H.J. de Vega }
\address{ Laboratoire de Physique Th\'eorique et Hautes Energies
     \foot{Laboratoire Associ\'e au CNRS UA 280}, Paris
     \foot{mail address: \nxl
           L.P.T.H.E., Tour 16 $1^{\rm er}$ \'etage, Universit\'e Paris VI,\nxl
           4 Place Jussieu, 75252, Paris cedex 05, France }}
\vfil
\abstract

The formulation and resolution of integrable lattice statistical
models in a quantum group covariant way is the subject of this review.
The Bethe Ansatz turns to be remarkably useful to implement
quantum group symmetries and to provide quantum group
representations even when $q$ is a root of unity.
We start by solving the six-vertex model with fixed boundary conditions (FBC)
that guarantee exact  $SU(2)_q$  invariance on the lattice.
The algebra of the Yang-Baxter (YB) and  $SU(2)_q$  generators turns to
close
and the transfer matrix is  $SU(2)_q$  invariant for FBC. In addition,
the infinite spectral parameter limit of the YB generators yields
{\bf cleanly}
the  $SU(2)_q$  generators. The Bethe Ansatz states constructed for FBC
are shown to be {\bf highest weights} of  $SU(2)_q$ .
The light-cone evolution
operator for FBC is introduced and shown to follow from the row-to-row
FBC transfer matrix with alternating inhomogeneities. This operator is
shown to describe the SOS model after an appropiate gauge choice. Using this
 FBC light-cone approach, the scaling limit of both six-vertex and SOS
models easily follows. Finally, the higher level Bethe Ansatz
equations
(describing the physical excitations) are explicitly derived for FBC.
We then solve the RSOS($p$) models on the light--cone lattice with fixed
boundary conditions by disentangling the type II representations of
$SU(2)_q$, at $q=e^{i\pi/p}$, from the full SOS spectrum obtained through
Algebraic Bethe Ansatz. The rule which realizes the quantum group reduction to
the RSOS states is that
there must not be {\bf singular} roots in the solutions of the  Bethe Ansatz
equations describing the states with quantum spin $J<(p-1)/2$. By studying how
this rule is active on the particle states, we are able to give a
microscopic derivation of the lattice $S-$matrix of the massive kinks.
The correspondence between the light--cone
Six--Vertex model and the Sine--Gordon field theory implies that the continuum
limit of the RSOS($p+1$) model is to be identified with the $p-$restricted
Sine--Gordon field theory.

\endpage

\REF\rev{H.J. de Vega, \IMPA{4}, 2371 (1989) and \nxl
 \NPB{\vphantom{0}}, {\sl Proc. Suppl.} {\bf 18A} (1990)
        229	. }
\REF\grq{V. Drinfeld, Proc. Int. Conf. Math. Berkeley (1986).
	L. D. Faddeev, N. Yu. Reshetikhin and L. A. Takhtadzhyan,
{\sl Algebra and Analysis}, {\bf 1}, 129 (1989).}
\REF\ddvI{C. Destri and H.J. de Vega, \NPB{374}(1992)692}
\REF\nos{C. Destri and H. J. de Vega \NPB{385} , 361 (1992) .}
\REF\qru{G. Lusztig. {\sl Contemp. Math.} {\bf 82} (1989) 59. \nxl
          P. Roche and D. Arnaudon, \LMP{17} (1989) 295.}
\REF\qrud{V. Pasquier and H. Saleur, \NPB{330}, 523 (1990).}
\REF\ppk{P. P. Kulish and E.K. Sklyanin, \JPA{24}, L435 (1991).}
\REF\euge{E.K. Sklyanin, \JPA{21}, 2375 (1988).}
\REF\ivan{I. Cherednik, \TMP{61}, 35 (1984).}
\REF\sing{H. J. de Vega, \IMPA{5}, 1611 (1990).}
\REF\ale{H. J. de Vega and A. Gonz\'alez Ruiz, LPTHE preprint
92-45 Newton Institute preprint 92012, to appear in \JPA{}}
\REF\aust{C. J. Hamer and M. T. Batchelor, \JPA{21} L173 (1988),\nxl
C. J. Hamer, G.R.W. Quispel and M. T. Batchelor, \nxl \JPA{20}, 5677 (1987).}
\REF\abf{G.E. Andrews, R. J. Baxter and P. J. Forrester,
\nxl \JSP{35}, 193 (1984).}
\REF\nons{F. Woynarovich, \JPA{15}, 2985 (1982).\nxl	C. Destri and
J. H. Lowenstein, \NPB{205}, 369 (1982) \nxl
O. Babelon, H. J. de Vega and C. M. Viallet, \NPB{220}, 13 (1983).}
\REF\asy{H.J. de Vega, \NPB{240}, 495 (1984).}
\REF\chico{F.C. Alcaraz et al., \JPA{20}, 6397 (1987).}
\REF\leon{ L. D. Faddeev and L. A. Takhtadzhyan, {\sl J. Sov. Math.} {\bf 109},
134 (1984).}
\REF\maww{L. Mezincescu and R.I. Nepomechie, \MPL{A6},
2497 (1991).}
\REF\ddv{C. Destri and H.J. de Vega, \NPB{290} (1987) 363.}
\REF\ddvb{C. Destri and H. J. de Vega,  \JPA{22}, 1329 (1989).}
\REF\many{A. LeClair, \PLB{230} (1989) 282. \nxl
          N. Reshetikhin and F. Smirnov, \CMP{131} (1990) 157. }

\baselineskip=14 true pt
\vsize=8.5 true in
\hsize=6 true in

\chapter{Introduction}

As is by now well known integrability is a
consequence of the Yang-Baxter equation (YBE) in two-dimensional
lattice models and two-dimensional quantum field theory (QFT),
(for recent reviews see
for example ref.[\rev ]) . More precisely, a statistical model is
integrable
when the local weights are solutions of the YBE. Analogously, for two
dimensional integrable QFT the two-body S-matrix fulfils the YBE.

Quantum groups are closely related to Yang-Baxter algebras[\rev ,\grq ].
However,
quantum group invariance holds for an integrable lattice model
only for specific choices of the boundary conditions. As we showed
in refs.[\ddvI -\nos]  (see also refs.[\qru,\qrud,\ppk]),
choosing fixed boundary
conditions (FBC), the transfer matrix commutes with the quantum group
generators.

The purpose of this paper
 is to review  the work in collaboration with Claudio Destri in
 refs.[\ddvI -\nos] on integrable lattice models and  their scaling
limit using a fully quantum group covariant Bethe Ansatz (BA)
framework. Let us recall that the YB algebra of monodromy operators
acting on the space of physical states, is the main tool to
construct the transfer matrix eigenvectors by the (algebraic)
Bethe Ansatz. To do that we choose FBC (Dirichlet type) and use
the Sklyanin-Cherednik[\euge ,\ivan ] construction of the Yang-Baxter algebra.
In this framework, besides the R-matrix defining the local statistical
weights, there are two matrices  $K^{\pm}(\t)$  that define the boundary
conditions.  $K^{\pm}(\t)$  must fulfil a set of equations [eqs.(2.12) and
(2.15)] in order to respect integrability. (FBC is one special case
out of a continuous family of boundary conditions compatible with
integrability).

The appropiate monodromy operators  $U_{ab}(\l,{\tilde \omega})$
for a N-sites line take now the form depicted in fig.1 where
arbitrary inhomogeneities $ \omega_i \; ,   (1 \leq  i \leq N)$,
 are allowed at each
site.	We compute the large  $\t$  limit of the monodromy operators
  $U_{ab}(\l,{\tilde \omega})$  and find that they are just
the quantum group generators.
We do that explicitly for the six-vertex model where the quantum
group is  $SU(2)_q$  . We find in this way an explicit representation
of the  $SU(2)_q$  generators acting on the space of states.

In addition, the  $\t \to \infty$   limit of the Yang-Baxter algebra
for the  $U_{ab}(\l,{\tilde \omega})$  shows that the transfer matrix
    $t(\l,{\tilde \omega})$
commutes with the  $SU(2)_q$  generators and that the algebra of the
 $U_{ab}(\l,{\tilde \omega})~ (1 \leq  a, b  \leq 2)$ with the
quantum group generators
( $J_{\pm}$  and  $q^{J_z}$ ) closes [see eqs.(2.33)-(2.37)]. Moreover,
  $t(\l,{\tilde \omega})$ in the   $\t \to \infty$   limit yields
the q-Casimir operator ${\cal C}_q$  through
$$
t(\infty ,{\tilde \omega}) = q + q^{-1} +  ( q - q^{-1})^2 ~ {\cal C}_q
\eqn\prim
$$
It must be recalled
that for periodic boundary conditions (PBC)  the   $\t \to \infty$   limit
of the Yang-Baxter algebra also gives the quantum group generators
but that the algebra with the PBC monodromy operators does not close
[\rev ] .

	Then, we investigate the BA construction in
this quantum group covariant framework. It is natural to define for
FBC a creation operator of pseudoparticles  ${\B}(\t)$ (proportional to
 $U_{12}(\l,{\tilde \omega})$ ) which is odd in $\t$  [eq.(3.3)].
 The exact eigenvectors
of the transfer matrix   $t(\l,{\tilde \omega})$   are then given by
$$
\Psi (\vec{v}) =  {\B}(v_1 ) {\B}(v_2 ).... {\B}(v_r )
\Omega
\eqn\segu
$$
where  $v_1, v_2, \ldots , v_r$ ,  fulfil the BA equations (3.4) (BAE)
and  $\Omega$  is the ferromagnetic ground state (3.2). We show
that only BAE roots with
strictly positive real part must be consider ($\Re v_j > 0$).
In particular,  $ v_j = 0 $  and purely imaginary roots must be
discarded. We map the BAE (3.4) for FBC onto BAE for PBC in 2N sites
[see eq.(5.2)]. We find the usual PBC BAE plus an extra source and
two important constraints :

a) the total number of roots is even and they are
symmetrically distributed with respect to the origin,

b) a root at
the origin as well as purely imaginary roots are excluded.

Therefore,
the antiferroelectric ground state (and the excitations on the top of
it)
contain always a hole at the origin. This hole combined with the
extra source accounts for the surface energy (see for example
ref.[\aust ]).

	Starting from the general BA state (1.2) we prove that
they are highest weights for  $SU(2)_q$ . That is,
$$
J_+ \Psi (\vec{v}) = 0
\eqn\terc
$$
provided the BAE (3.4) hold for   $v_1, v_2, \ldots , v_r$ .
Therefore, the eigenvalues of   $t(\t,{\tilde \omega})$  are degenerate with
respect to the quantum group and the eigenvectors :
$$
J_- \Psi (\vec{v}),\; (J_-)^2 \Psi (\vec{v}),\ldots ,\;(J_-)^{2J} \Psi
(\vec{v}),
\eqn\falt
$$
are linearly independent from  $\Psi (\vec{v})$ .

SOS and vertex models are related by the vertex-face correspondence.
(The degrees of freedom lie on faces for SOS models whereas they
lie on links for the vertex model). The correspondence between them
 amounts to an application of the $q$-analog of the Wigner-Eckart theorem.
As we explain in sec. 4,
the SOS space of states is {\bf identical} to the set of maximal weight six
vertex states in a  quantum group invariant framework like ours.
It then follows from eq.\terc\ that the Solid--On--Solid (SOS)
states {\bf are just} the
f.b.c. BA states given by eq.\segu . The six vertex Hilbert space
includes the whole
$SU(2)_q$ multiplets and follow then by repeatedly applying the lowering
operator $J_-$ to the highest weight BA states \segu\ as shown in
eq.\falt .
In other words, we have found the BA solution of the SOS model
since we derived  the transfer matrix eigenstates.
In ref.[\sing ] an alternative but equivalent solution of the
SOS model is derived using  a BA in {\bf face} language. In addition
PSOS (periodic SOS models where the face states ${\it l}$ and
${\it l+p}$ are identified) are solved in ref.[\sing ].

The light-cone approach is a direct way to give a field theoretical
interpretation to a lattice model and furthermore obtain its
scaling limit as a massive QFT. The light-cone approach with
periodic boundary conditions has been investigated in refs.[\ddv
-\ddvb ] .
Here we consider this approach in the case of fixed boundary
conditions leading to a quantum group invariant framework.

In this context we show in sec. 4 that the diagonal-to-diagonal
transfer matrix  $U(\Th)$ [fig.6] can be obtained from the row-to-row
transfer matrix   $t(\l,{\tilde \omega})$  choosing the
inhomogeneities appropiately.
This is analogous to the relation found in ref.[\ddv ,\ddvb ] for PBC.
{}From the transfer matrix  $U(\Th)$  we define the lattice hamiltonian
through
$$
H = {i \o a}\log U(\Th)
\eqn\cuar
$$
where  $a$  is the  lattice spacing. In the
  $a \to  0 , \Th \to \infty$  limit this operator defines the continuous
QFT hamiltonian provided the renormalized mass scale [see eq.(4.17)]
is kept fixed.
	The evolution operator    $U(\Th)$  can be
considered both in the vertex or in the face language. In face
language, it has a simple expression provided we make an appropiate
gauge transformation. That is, if one transforms each local
$R$-matrix to a matrix ${\tilde R}$ such that the $SU(2)_q$ symmetry holds
locally. In this way, we show that the associated light-cone
 evolution operator   ${\tilde U}(\Th)$  just describes the SOS ABF
model[\abf ] [eq.(4.27)].

	As it is known, the physical states above the AF vacuum are
described by
the higher level BAE[\nons ]. We obtain the higher level BAE for FBC
in sec. 5 [eq.(5.14)]. They are the starting point to study
the excitations in SOS and RSOS models . To conclude we show
that the BAE for FBC admit solutions at infinity only when
$\g/\pi$  is a rational number. This is precisely the case when
 RSOS models can be defined and when the representations of
$SU(2)_q$ algebras cease to be isomorphic to usual SU(2). Recall that when
$\g/\pi$  is rational, type I representations are reducible but
indecomposable, type II are irreducible as in SU(2) (see for
example ref.[\qru ,\qrud ]).

It is known that the Six--Vertex (6V) model, in the so--called light--cone
formulation and with periodic boundary conditions (p.b.c.), yields the
Sine--Gordon massive field theory in an appropriate scaling limit [\ddv]. Hence
the light--cone 6V model can be regarded as an exactly integrable lattice
(minkowskian) regularization of the SG model.

Recently, the hidden invariance of the SG model under the quantum group
$SU(2)_q$ was exhibited [\many].
Our quantum group invariant light-cone  formulation,
provides a lattice formulation where such hidden invariance
appears starting from first principles. That is,
not on a bootstrap framework but deriving the field theory as
a rigorous scaling limit of the six-vertex model.

We present in section 6 and 7 the
derivation of the factorized $S-$matrices {\it on the lattice}, \ie\
still in the presence of the UV cutoff. This derivation is based on the
``renormalization" of the BA Equations, which consists in removing the
infinitely many roots describing the ground state. What is left is once again
a f.b.c. BA structure involving the lattice rapidities of the physical
excitations (the particles of the model) and the roots of the higher--level
BAE obtained in paper I. The explicit form of two--body $S-$matrix for the 6V
model and the SOS model can be extracted in a precise way (cft. eq.
(6.11)) from this higher--level BA structure.
In the massive scaling limit these
lattice scattering amplitudes become the relativistic $S-$matrices of the SG
model (or Massive Thirring model) and of the continuum SOS model.
Let us remark that the SOS $S-$matrix, although closely related to the 6V and
SG $S-$matrices from the analytical point of view, is conceptually different.
It describes the scattering of {\it kinks} interpolating between
{\it renormalized} local vacua labelled by integers. This kink $S-$matrix is
most conveniently expressed in the so--called face language (see eq.(7.3)).

In section 8 we investigate the f.b.c. BAE (eq.(2.4)) when the quantum group
deformation parameter $q$ is a root of unity, say $q^p=\pm 1$, with $p$ some
integer larger than 2 (the case $p=2$ being trivial). In this case it is known
that RSOS($p$) model can be introduced by restricting to the finite set
$(1,2,\ldots,p-1)$ the allowed values of the local height variables of the
SOS model [\abf].
At the same time  when $q^p=\pm 1$, the representations of
$SU(2)_q$ algebras cease to be isomorphic to usual SU(2). Recall that when
$\g/\pi$  is rational, type I representations are reducible but
indecomposable, type II are irreducible as in SU(2) (see for
example ref.[\qru ,\qrud ]).
The restriction leading to the RSOS model from the SOS model
 is equivalent to the projection of the full SOS Hilbert
space (which is formed by the highest weight states of $SU(2)_q$) to the
subspace spanned by the type II representations [\qru ,\qrud ]. That is, those
representations which remain irreducible when $q$ becomes a root of unity. Our
results on this matters, in the BA context, can be summarized as follows:
\item{a)} Only when $q$ is a root of unity, the f.b.c. BAE admit singular roots
          (that is vanishing $z-$roots or diverging $v-$roots, in the
          notations of sec.1).
\item{b)} When $q$ tend to a root of unity, say $q^p=\pm1$, the BAE
          solutions  can be divided
          into regular and singular solutions, having, respectively, no
          singular roots or some singular roots. Regular solutions correspond
          to irreducible type II representations. Singular solutions with
          $r$ singular roots correspond to the reducible and generally
          indecomposable type I representations obtained by mixing two standard
          $SU(2)_q$ irreps of spin $J$ and $J+r$ (we recall that $J=N/2-M$,
          where $N$ is the spatial size of the lattice and $M$ the number
          of BA roots). Then necessarily $r<p$.
\item{c)} The $r$ singular roots $z_1,z_2,\ldots,z_r$ vanish with fixed ratios
          $$
                  z_j=\omega^{j-1}z_1  \;,\quad \omega=e^{2\pi i/r}
                      \;,\quad 1\le j\le r                        \eqn\first
          $$
          In terms of the more traditional hyperbolic parametrization, with
          $v-$roots related by $v_j=-\frac12\log z_j$ to the $z-$roots, the
          singular roots form an asymptotic string--like configuration. They
          have a common diverging real part and are separated by $\pi/r$ in
          the imaginary direction.

So we see that the RSOS eigenstates are easily singled out from the
full set of BA eigenstates of the 6V or SOS transfer matrix. One must retain
{\bf all and only those BAE solutions with} ${\bf M>(N-p+1)/2}$
(which correspond to states
with $J<(p-1)/2$) {\bf and with all }${\bf M~ z-}${\bf roots different from
zero}.  This provides
therefore an exact, explicit and quite simple solution for the RSOS model on
the lattice (with suitable boundary conditions, as we shall later see).
In particular the ground state of the 6V, SOS and RSOS models is the
same f.b.c. BA state (in the thermodynamic limit
$N\to\infty$ at fixed lattice spacing).
It is the unique $SU(2)_q$ singlet with all real positive roots and no holes.
The local height configuration which, loosely speaking, dominates this ground
state can be depicted as a sequence of bare kinks jumping back and forth
between neighboring bare vacua (see fig. 10). In the massive scaling limit
proper of the light--cone approach [\ddv], this ground state becomes the
physical vacuum of the SG model as well as of and all the Restricted SG field
theories.

In the discussion ending sec. 8, we argue that the kink $S-$matrix for the
excitations of the RSOS models follows indeed by restriction from that of the
SOS model. In the scaling limit it is to be identified with the relativistic
$S-$matrix of the Restricted SG models. All these field--theoretical
$S-$matrices are naturally related to the  Boltzmann weights of the respective
lattice models. This is because the  higher--level BAE are identical in form to
the ``bare" BAE, apart from the renormalization of the anisotropy parameter
$\g$ (related to $q$ by $q=e^{i\g}$)
$$
          \g\to {\hat \g}={{\pi\g}\o{\pi-\g}}                     \eqn\iren
$$
and the replacement of the rapidity cutoff $\pm\Th$ with the suitably scaled
rapidities  $\t_j$ of the physical excitations
$$
       \pm\Th \to  {{\g\t_j}\o{\pi-\g}}                           \eqn\repl
$$
In particular, since $\g=\pi/p$ for the RSOS($p$) model, eq.\iren\ yields
$p\to p-1$. This shows that the renormalized local vacua ${\hat\ell}$ run from
1 to $p-2$ when the bare local heights $\ell_n$ run from 1 to $p-1$. The
higher--level BA structure of the light--cone 6V model (or lattice regularized
SG model) thus provides a microscopic derivation of the bootstrap construction
of ref. [\many] and  explains why the $S-$matrix of the $p-$restricted SG field
theory has the same functional form of the Boltzmann weights of the lattice
RSOS($p+1$) model. Moreover, the well--known correspondence between the
critical RSOS($p$) models and the Minimal CFT Models $M_p$ imply the natural
identification of the massive $p-$restricted SG model with a completely
massive relevant perturbation of $M_p$. This is generally recognized as the
perturbation induced by the primary operator $\phi_{1,3}$ with negative
coupling.

Two detailed examples of the BA realization of the quantun group reduction to
the RSOS models are presented in section 9. We considered the simplest cases
$p=3$ and $p=4$.
The RSOS(3) model is a trivial statistical system with only one state, since
all the local heigths are fixed once we choose, for example, the boundary
condition $\ell_0=1$. In our quantum group covariant f.b.c. construction
this corresponds to the existence of one and only one type II state when
$\g=\pi/3$. We then obtain the following purely mathematical result:
for any $N\ge2$ and real $w=\exp(-2\Th)$ the set of BAE
$$
     \left({{z_j w-e^{\pi i/3}}\o{z_j we^{\pi i/3}-1}}
     \;{{z_j-we^{\pi i/3}}\o{z_j e^{\pi i/3}-w}}\right)^N =
     \prod_{\scriptstyle k=1\atop\scriptstyle k\ne j}^{[N/2]}
     {{z_j-z_k e^{2\pi i/3}}\o{z_j e^{2\pi i/3}-z_k}}
     \;{{z_j z_k-e^{2\pi i/3}}\o{z_j z_k e^{2\pi i/3}-1}}
            \;,\quad 1\le j\le N                                 \eqn\baetre
$$
admit one and only one solution with non--zero roots within the unit disk
$|z|<1$. In addition, these roots are all real and positive.

The RSOS(4) model can be exactly mapped into an anisotropic Ising model as
showed in eqs. (9.1--5). In our case the horizontal and vertical Ising
couplings turn out to be $\Th-$dependent {\it complex} numbers. For even $N$,
the Ising spins are fixed on both space boundaries. For odd $N$ the spins
are fixed on the left and free to vary on the right. We analyzed the BAE for
the RSOS(4) model in some detail. In the thermodynamic limit the ground state,
as already stated, is common to the 6V and SOS moodels. The elementary
excitations correspond to the presence of holes in the sea of real roots
charactering the ground state. Each holes describes a physical particle or kink
and may be accompanied by complex roots. In sec. 9 we argue that in the RSOS(4)
case a state with $\nu$ holes necessarily contains $[\nu/2]$ {\it two--strings}
those position is entirely fixed by the holes. Notice that here the number
$\nu$ of holes can be odd even for even $N$. This is not the case for
the 6V or SOS models, where $\nu$ is always even for $N$ even. What happens is
that when $\g\to(\pi/4)^-$ the largest real $v-$root diverges in the $J=1$
states of the 6V and SOS models. Therefore, these states get mixed with $J=2$
states into type I representations and do not belong to the RSOS(4) Hilbert
space. The RSOS(4) states with $J=1$ are obtained by choosing the largest
quantum integer $I_{N/2-1}=N/2$ (see the Appendix). There is no root associated
to $N/2+1$. It follows that these states, from the 6V and SOS viewpoint, have a
cutoff dependent term in the  energy equal to $\pi/a$, where $a$ is the lattice
spacing (loosely speaking, one could say that ``there is a hole at infinity").
They are removed from the physical SG spectrum in the continuum limit. We are
thus led to propose as RSOS(4) hamiltonian, for even $N$
$$
          H_{RSOS(4)}=H_{SOS}(\g=\pi/4)- a^{-1}\pi J           \eqn\hamil
$$
where $H_{SOS}$ is given by eqs. (4.14), (4.10), (4.21-27) and $J=0$ or 1.
In this way, the particle content of the light--cone RSOS(4) and the
corresponding $S$-matrix turn out to coincide with the results
of the bootstrap--like approach of ref. [\many]. Eq. \hamil\ defines, in the
scaling limit, the hamiltonian of the $(p=3)-$restricted SG model.
Notice, in this respect, that the higher--level BAE (5.17) and (6.10)
completely determine the physical states in terms of renormalized
parameters.

To summarize, the picture we get from the BA solution of the f.b.c. 6V, SOS and
RSOS($p$) lattice models is as follows. Performing the scaling limit whithin
the light--cone approach, these lattice models yields respectively:  the SG
model (or Massive Thirring Model), a truncated SG and the $(p-1)$restricted SG
models. For the SG model we have essentially nothing to add to the existing
literature, apart from the explicit unveiling, at the regularized  microscopic
level, of its $SU(2)_q$ invariance and for a better derivation of the
$S-$matrix. The truncated SG follows by keeping only the highest weight states
with respect to $SU(2)_q$, that is the kernel of the raising operator $J_+$.
Finally we showed that the RSOS($p$) lattice models with trigonometric weights
yield in the scaling limit proper of the light--cone approach the
$(p-1)$restricted SG field theories formulated at the bootstrap
level in ref. [\many].

\chapter{Boundary conditions in lattice integrable models}

Let us consider an integrable vertex model with R-matrix
$R^{ab}_{cd}(\t)$  (see fig.1). Each element $R^{ab}_{cd}(\t)$
defines the statistical
weight of this configuration. We assume  $R(\t)$  to fulfil the
Yang-Baxter equations. The indices $a, b, c, d,$ are assumed to
run from  1  to  $q$  with  $q \ge 2$.
$$
\left[ {\bf 1} \otimes R(\t - \t') \right]\;
\left[  R(\t) \otimes {\bf 1} \right]\;
\left[ {\bf 1} \otimes R(\t') \right]\;
=
\left[  R(\t') \otimes {\bf 1} \right]\;
\left[ {\bf 1} \otimes R(\t) \right] \;
\left[  R(\t - \t') \otimes {\bf 1} \right]
\eqn\ybe
$$
We shall assume  T  and  P  invariance for  $R(\t)$
$$
R^{ab}_{cd}(\t) =R^{cd}_{ab}(\t)=R^{ba}_{dc}(\t)
\eqn\invpt
$$
In addition, we assume  $R(\t)$  to be regular,
 that is
$$
 R(0) = c~ {\bf 1} \qquad {\rm or}\qquad R^{ab}_{cd}(0) = c ~\delta^a_c
\delta^b_d
\eqn\regu
$$
where $c$ is a numerical constant. Eqs. \ybe\ and \regu\  imply the
unitarity-related equation
$$
 R(\t)\; R(-\t)= \rho(\t)\;  {\bf 1}
\eqn\unit
$$
where $\rho(\t)$ is an even c-number
function.
	Furthermore, we assume crossed-unitarity.
That is,
$$
 {\hat R}(\t)\; {\hat R}(-\t - 2 \eta)= {\hat \rho}(\t)\;  {\bf 1}
\eqn\crun
$$
where  $\eta$  is a constant and
${\hat R}^{ab}_{cd}(\t) \equiv R^{ac}_{bd}(\t)$ .

Let us consider now a NxN' square lattice with periodic boundary
conditions. Then the row-to-row transfer matrix is given by
$$
\tau(\t,{\tilde \omega})=\sum_a T_{aa}(\t,{\tilde \omega})
\eqn\trans
$$
where the operators  $T_{ab}(\t,{\tilde \omega})$  defined by (see
fig.2)
$$
T_{ab}(\t,{\tilde \omega})= \sum_{a_1,...,a_{N-1}} t_{a_1b}(\t -
\omega_1 ) \otimes  t_{a_2a_1}(\t - \omega_2 )\otimes .....
\otimes  t_{aa_{N-1}}(\t - \omega_N )
\eqn\monod
$$
act on the vertical space ${\cal  V}  = \bigotimes_{1 \leq i \leq N}
V_i ~~, V_i \equiv  C^q$
 , and the simplest choice for the local vertices is $\left[
t_{ab}(\t)\right]_{cd} \equiv R^{bd}_{ca}(\t) $.
 In eq.\monod\  $ \omega_1 , \omega_2 ,...., \omega_N $
stand for arbitrary inhomogeneity parameters.
The  $T_{ab}(\t,{\tilde \omega})$
fulfil the YB algebra
$$
 R(\l-\mu)\left[T(\l,{\tilde \omega} )\otimes
     T(\mu,{\tilde \omega} )\right] =\left[T(\mu,{\tilde \omega} )\otimes
     T(\l,{\tilde \omega} )\right]R(\l-\mu)
 \eqn\nuda
$$
as follows from eq.\ybe . Thus, the  $\tau(\t,{\tilde \omega})$
  are a commuting family
$$
\left[\tau(\t,{\tilde \omega}) ,\tau(\t',{\tilde \omega})\right] = 0
\eqn\comut
$$
Let us now consider the
generalization to other boundary conditions compatible with
integrability[\euge]. Define (see fig.3)
$$
U_{ab}(\t,{\tilde \omega}) = \sum_{cd}T_{ad}(\t,{\tilde \omega})
K^-_{dc}(\t) T_{cb}^{-1}(-\t,{\tilde \omega})
\eqn\matru
$$
where  $ T^{-1}(-\t,{\tilde \omega})$  is the inverse in both the horizontal
and vertical spaces :
$$
\sum_b T_{ab}(\t,{\tilde \omega}) T_{bc}^{-1}(-\t,{\tilde \omega})
=  ~\delta^a_c ~{\cal I}
\eqn\tinver
$$
and  $ {\cal I}$ is the identity on the vertical space   ${\cal  V}$ .
Summation over indices of the vertical space ${\cal  V}$ are omitted
both in eqs.\matru and \tinver  (cft. fig. 3).
$K^-(\t)$ in eq.\matru\  is a  qxq  matrix
solely acting on the horizontal space. It must fulfil[\euge]
$$\eqalign{
R(\l-\mu)\left[ K^-(\l) \otimes {\bf 1}  \right] & R(\l+\mu)\;
\left[ K^-(\mu) \otimes {\bf 1} \right] = \cr
& \left[ K^-(\mu) \otimes {\bf 1} \right] R(\l+\mu)
\left[ K^-(\l) \otimes {\bf 1} \right]
R(\l-\mu) \cr}
\eqn\matk
$$
(Notice, that our R-matrix differs from ref.[\euge] in a
permutation matrix  $R \to P R ,  \nxl P^{ab}_{cd}(\t) = ~\delta^a_d
\delta^b_c )$.

As is well known,  $R^{ab}_{cd}(\t - \t')$   has the interpretation of
scattering
amplitude for a two-body collision where $a(d)$ and $b(c)$ label the
 initial (final) states of two particles with rapidities  $\t$  and
 $\t'$  respectively. In this S-matrix context,  $K^-_{ab}(\t)$  is the
scattering amplitude for one particle with a rigid wall on the
left, a and b labelling the initial and final states and $\t$ being
its final rapidity (see fig.4).

	Thanks to eqs.\nuda\  and \matk , $U(\t,{\tilde \omega})$
fulfils the Yang-Baxter algebra
$$\eqalign{
R(\l-\mu)\left[ U(\l,{\tilde \omega})\otimes {\cal I} \right]&R(\l+\mu)
\left[ U(\mu,{\tilde \omega}) \otimes {\cal I} \right] = \cr
& \left[ U(\mu,{\tilde \omega}) \otimes {\cal I} \right]R(\l+\mu)
\left[  U(\l,{\tilde \omega}) \otimes {\cal I} \right]
R(\l-\mu) \cr}
\eqn\ybau
$$
The transfer matrix is given now by
$$
t(\l,{\tilde \omega}) = \sum_{ab} K^+_{ab}(\l + \eta)U_{ab}(\l,{\tilde \omega})
\eqn\tranu
$$
Here  $K^+(\l)$  describes the scattering with a rigid wall on the
right (cft. fig.5). It is a solution of the equation :
$$\eqalign{
R(\l-\mu)\left[ {\bf 1} \otimes K^+(\l)  \right] & R(\l+\mu)
\left[  {\bf 1}\otimes  K^+(\mu) \right] = \cr
& \left[ {\bf 1} \otimes  K^+(\mu) \right] R(\l+\mu)
\left[ {\bf 1} \otimes K^+(\l) \right]
R(\l-\mu) \cr}
\eqn\matkp
$$
It follows from eqs.\ybau -\matkp\  that $t(\l,{\tilde \omega})$ is a
commuting family
$$
\left[\;t(\t,{\tilde \omega}) ,\;t(\t',{\tilde \omega})\right] = 0
\eqn\comut
$$
As we can see from eqs. \matru\ and \tranu\ the
boundary conditions associated to $t(\t,{\tilde \omega})$
 follow from the form of $K^+(\l)$ and   $K^-(\l)$ sitting on
the right and left borders, respectively.

{}From now on we shall consider the case of the six-vertex model, where
q = 2  and the  R--matrix reads, in terms of a generic
spectral parameter $\t$,
$$  \eqalign{R(\t)=\pmatrix{1&0&0&0\cr
                    0&c&b&0\cr
                    0&b&c&0\cr
                    0&0&0&1\cr} \cr}     \qquad
    \eqalign{b=&\;b(\t,\g)={{\sinh\t}\o{\sinh(i\g-\t)}} \cr
             c=&\;c(\t,\g)={{\sinh i\g}\o{\sinh(i\g-\t)}}}
\eqn\rma
$$
where the anisotropy parameter $\g$ (we may assume $0\le\g\le\pi$) is related
to
the quantum group deformation $q$ by $q=\exp(i\g)$. This R--matrix
\rma\ is unitary for real $\t$.
Crossed unitarity \crun\  holds for  $\eta  = i\g$ . It is also convenient
to work with slightly modified local vertices
$$
\left[ t_{ab}(\t)\right]_{cd} \equiv R^{bd}_{ca}(\t - i \g /2)
\eqn\mver
$$
in order to construct the row-to-row monodromy
matrix  $T_{ab}(\t,{\tilde \omega})$ . For this R-matrix the diagonal solutions
$K^{\pm}(\t)$ of eqs.\matk\  and \matkp\  turn to be[\euge,\ivan]
$$
K^{\pm}(\t) = K(\t,\xi_{\pm} )
$$
where
$$
 \eqalign{K(\t, \xi)={1 \o {\sinh{\xi}}} \pmatrix{\sinh(\xi+\t)&0\cr
                    0&\sinh(\xi - \t)\cr} \cr}
\eqn\soldi
$$
Here $\xi_+$ and  $\xi_-$ are arbitrary numbers that parametrize
this  boundary conditions compatible with integrability.
(For the general solution of eqs.\matk\  and \matkp\ in the six-vertex
model see ref.[\ale ].)

We are interested on boundary conditions yielding a quantum
group covariant framework. [$SU(2)_q$ for the six-vertex model].
For periodic boundary conditions the quantum group transformation
 properties of $T_{ab}(\t,{\tilde \omega})$ and $t(\t,{\tilde
\omega})$  are not simple [\asy].
A $SU(2)_q$ invariant XXZ hamiltonian requires fixed boundary
conditions and special end-point terms[\qrud ].

The XXZ hamiltonian follows from  $dt(\t, {\tilde \omega} = 0)/d\t$ evaluated
 at  $\t = 0$ . One finds
 from eqs.\tranu ,\rma , and \soldi\  [\euge]
$$\eqalign{
H_{XXZ} =& - { 1 \o { 4 trK^+(0)}} {d \o {d\t}} \left[
t(\t, {\tilde \omega} = 0) - tr K^+(\t) \right] \cr =&
- {1 \o 2} \sum_{n=1}^{N-1}(\s_n^x \s_{n+1}^x +
\s_n^y \s_{n+1}^y + \cos\g \s_n^z \s_{n+1}^z )  \cr
& + {1\o 2} \sinh(i\g)(\s_1^z\,\coth\xi_- + \s_N^z\,\coth\xi_+ ).\cr}
\eqn\xxz
$$
As one can check, the quantum group invariant
 case corresponds to  $\xi_{\pm}= \pm \infty$.
	We choose therefore   $\xi_{\pm}= \pm \infty$  in order to
built a $SU(2)_q$
covariant framework for the six-vertex model and its scaling
limit. This choice corresponds physically to boundary conditions
 of Dirichlet type. In that case
$$
 \eqalign{K^{\pm}(\t)=\exp(\mp\t\s_z) = \pmatrix{e^{\mp \t}&0\cr
                    0&e^{\pm\t}\cr} \cr}
\eqn\soliq
$$
Now, in order to find the $SU(2)_q$ content of
the YB algebra \tranu\  and its associate Bethe Ansatz
construction, we start by computing the $\t \to \infty$ limit of
the  $U_{ab}(\t,{\tilde \omega})$ operators.

The   $\t \to \pm \infty$  limit of the row-to-row monodromy matrix
$$\eqalign{
T(\t,{\tilde \omega}) = \pmatrix{A(\t)&B(\t)\cr
                    C(\t)&D(\t)\cr} \cr}
\eqn\abcd
$$
yields  $SU(2)_q$  generators [\rev ,\asy ]. We have
$$\eqalign{
A(\t)\buildrel{\t\to \pm\infty}\over =& e^{\mp iN\g/2}~\exp(\pm i \g J_z
)~ [ 1 + O(e^{\mp 2\t})] ~~,\cr
D(\t)\buildrel{\t\to \pm\infty}\over =& e^{\mp iN\g/2}~\exp(\mp i \g J_z
)~ [ 1 + O(e^{\mp 2\t})] ~~,\cr
B(\t)\buildrel{\t\to \pm\infty}\over =& \pm 2 e^{\mp iN\g/2 \mp\t}~
\sinh(i\g) ~ J_-(\mp{\tilde \omega},\mp \g) ~ [ 1 + O(e^{\mp 2\t})] ~~,\cr
C(\t)\buildrel{\t\to \pm\infty}\over =& \pm 2 e^{\mp iN\g/2 \mp\t}~
\sinh(i\g) ~ J_+(\pm{\tilde \omega},\pm \g)~  [ 1 + O(e^{\mp 2\t})] ~~,\cr}
\eqn\asab
$$
where
$$
J_{\pm}({\tilde \omega},\g) = \sum_{k=1}^N \prod_{j=1}^{k-1}
\exp[-{{i\g}\o 2}(\s_j)_z ] e^{\pm \omega_k} (\s_{\pm})_k
\prod_{l=k+1}^N \exp[ {{i\g}\o 2}(\s_l)_z ]
\eqn\jpmq
$$
and
$$
J_z = {1 \o 2} \sum_{a=1}^N(\s_a)_z
\eqn\jotaz
$$
Notice that  $J_+ \equiv  J_+({\tilde \omega},\g) ,\;  J_- \equiv  J_-({\tilde
\omega},\g) $  and  $ J_z $  are $SU(2)_q$ generators with  $q =
e^{i\g}$ , for they obey the commutation
rules
$$\eqalign{
[ J_+ , J_- ] =& \; {{\sin(2\g J_z )} \o {\sin \g}} =
 {{q^{2 J_z} -q^{-2 J_z}} \o { q - q^{-1}}} \cr
[ J_z , J_{\pm} ] =& \pm J_{\pm} \cr}
\eqn\grupq
$$
For the boundary conditions   $\xi_{\pm}= \pm \infty$ , one sets
$$\eqalign{
U(\t,{\tilde \omega}) = \pmatrix{{\cal A}(\t)&{\cal B}(\t)\cr
                    {\cal C}(\t)&{\cal D}(\t)\cr} \cr}
\eqn\uabc
$$
{}From eqs.\matru\  and \soliq\  it follows that
$$\eqalign{
{\cal A}(\t) =& e^{i\g/2 - \t}~B(\t)~C(-\t)\;
- \; e^{-i\g/2 + \t}~A(\t)~D(-\t)\cr
{\cal B}(\t) =& e^{-i\g/2 + \t}~A(\t)~B(-\t)\;
- \; e^{i\g/2 - \t}~B(\t)~A(-\t)\cr
{\cal C}(\t) =& e^{i\g/2 - \t}~D(\t)~C(-\t)\;
- \; e^{-i\g/2 + \t}~C(\t)~D(-\t)\cr
{\cal D}(\t) =& e^{-i\g/2 + \t}~C(\t)~B(-\t)\;
- \; e^{i\g/2 - \t}~D(\t)~A(-\t)\cr }
\eqn\cala
$$
Let us compute the $\t \to \infty$  limit of these operators. We
find from eqs.\asab\ -\cala\
$$\eqalign{
{\cal A}(\t)\buildrel{\t\to \infty}\over =& - \; e^{-i\g/2 + \t}~
\exp(2i\g J_z )~ [ 1 + O(e^{-2\t})] ~~,\cr
{\cal B}(\t)\buildrel{\t\to \infty}\over =& - \; e^{-i\g/2}~
\sinh(i\g)~ J_-~ [ 1 + O(e^{-2\t})] ~~,\cr
{\cal C}(\t)\buildrel{\t\to \infty}\over =& - \; e^{-i\g/2}~
\sinh(i\g)~ J_+~e^{i\g J_z }~ [ 1 + O(e^{-2\t})] ~~,\cr
{\cal D}(\t)\buildrel{\t\to \infty}\over =&  \;
2 e^{-i\g/2-\t} \sin\left[(J+J_z)\g\right]
 \sin\left[(J-J_z + 1)\g\right] \cr &- {1 \o 2}e^{i\g/2}
\;e^{-2i\g J_z-\t } \;  + O(e^{-3\t}) ~~,\cr}
\eqn\ascal
$$
The operator  $J$  is defined through the $SU(2)_q$ Casimir
invariant  ${\cal C}_q$
$$
{\cal C}_q = {1 \o 2} ( J_+ J_- + J-- J_+ ) + \cos\g \;
{{\sin^2(\g J_z)} \o {\sin^2(\g)}} =
{{\sin \g (J+1) \sin \g J } \o {\sin^2 \g}}
\eqn\qcas
$$
As one can see from eqs.\ascal\  the asymptotic form of
 $U_{ab}(\l,{\tilde \omega})$
  is related with the  $SU(2)_q$  generators in a very clean way.

Let us now consider the transfer matrix  $t(\l,{\tilde \omega})$
  [eq.\tranu\ ].
We find when  $\xi_{\pm}= \pm \infty$
$$
t(\t,{\tilde \omega})= e^{-i\g/2 - \t}~{\cal A}(\t) +
 e^{i\g/2 + \t}~{\cal D}(\t)
\eqn\trunu
$$
Now, when $\t \to \infty$ ,  inserting eq.\ascal\  in \trunu\  yields
$$
t(\t,{\tilde \omega})\buildrel{\t\to \infty}\over =
2 \cos[\g(2J+1)] + O(e^{-2\t})
\eqn\asitu
$$
That is, the asymptotics of the transfer matrix expresses up to
numerical constants in terms of the q-Casimir ${\cal C}_q$  [eq.\qcas\ ].

The operators   $U_{ab}(\t,{\tilde \omega})$  fulfil the Yang-Baxter
algebra \ybau\ with R-matrix \rma\ .
Letting  $\t' \to \infty$  in eq.\ybau\  we can compute
the relevant commutation of   $U_{ab}(\t,{\tilde \omega})$
  with the  $SU(2)_q$   generators.
We find
$$
\left[ {\cal A}(\t) , J_- \right] = - {\cal B}(\t)~e^{i \g J_z + \t}\quad
 ' \quad \left[ {\cal D}(\t) , J_- \right] =  {\cal B}(\t)~e^{i\g( J_z +1)+ \t}
\eqn\cadj
$$
It is well known
that  $B(\t)$  and  $C(\t)$  act as lowering and raising operators
for  $J_z$ , while  $A(\t)$  and  $D(\t)$  commute with it[\rev].
We find here the same properties for the elements of  $U(\t,{\tilde
\omega})$,
$$\eqalign{
\left[ \; {\cal B}(\t) ,\; J_z \right] &=  {\cal B}(\t)~,~
\left[ \; {\cal C}(\t) ,\; J_z \right] =  -{\cal C}(\t)~, \cr
\left[ \; {\cal A}(\t) ,\; J_z \right] &=\left[ \; {\cal D}(\t) , J_z \right] =
\left[\; t(\t,{\tilde \omega})  ,\;  J_z \right]=0 \cr}
\eqn\ccaj
$$
In addition, we find
$$
\left[\;e^{-i\g J_z}\;  {\cal B}(\t) ,\; J_- \right] = 0~~,~~
 {\cal B}(\t)~J_+ =e^{-i\g}\; J_+ ~  {\cal B}(\t) +
[\; e^{\t - i \g}\; {\cal D}(\t) - e^{-\t} \;{\cal A}(\t) ] ~e^{i\g J_z}~
\eqn\bjma
$$
Using now eq.\trunu\  yields
$$
\left[\;t(\t,{\tilde \omega})  ,\; J_- \right]=0
\eqn\tjme
$$
One can analogously prove that
$$
\left[\; t(\t,{\tilde \omega})  ,\; J_+ \right]=0
\eqn\tjma
$$
Therefore, the transfer matrix  $t(\t,{\tilde \omega})$  is  $SU(2)_q$
  invariant. As a corollary, we see that  $H_{XXZ}$  (see eq.\xxz )
is $SU(2)_q$-invariant, as shown in ref.[\qrud ]  by direct
calculation.

We investigate in the next section the
Bethe Ansatz construction of eigenvectors of  $t(\t,{\tilde \omega})$.
The first consequence of the the  $SU(2)_q$
  invariance of  $t(\t,{\tilde \omega})$ is the degeneracy of its
eigenvalues   with respect to the quantum
group.

\chapter{The Bethe Ansatz and the $SU(2)_q$ group}

In section II we developped the Yang-Baxter framework
with boundary conditions adapted to the $SU(2)_q$ invariance.
We call this a quantum group covariant framework.
Let us now investigate in such scheme the eigenvectors of the transfer
matrix  $t(\t,{\tilde \omega})$    in the algebraic Bethe Ansatz[\euge ].
These eigenvectors can be written as
$$
\Psi (\vec{v}) =  {\B}(v_1 ) {\B}(v_2 ).... {\B}(v_r )
\Omega
\eqn\abal
$$
where $\Omega$ is the ferromagnetic ground state and $\vec{v} \equiv
(v_1 ,v_2 , \ldots ,v_r ) $.
$$
\Omega = \pmatrix{1 \cr 0 \cr} \otimes \pmatrix{1 \cr 0 \cr} \otimes
..... \otimes \pmatrix{1 \cr 0 \cr}
\eqn\refe
$$
and
$$
 {\B}(\t) \equiv {{\sinh2\t}\o{\sinh(2\t-i\g)}}~e^{i\g/2} \;
 {\cal B}(\t) = \;e^{- \t}~B(\t)~A(-\t)- e^{\t}~B(-\t)~A(\t)
\eqn\bsom
$$
Notice that use has
been made of the Yang-Baxter algebra \nuda\  to obtain
the last form in eq.\bsom\  from the expression \cala\  for $ {\cal B}(\t)$.
The numbers  $v_1, v_2, ...., v_r ,   (0 \le r \le  N/2)$
 must be all distinct roots of the set of algebraic
equations
$$\eqalign{
&\prod_{k=1}^N
     {{\sinh[v_m - \omega_k+i\g/2]}\o{\sinh[v_m -\omega_k-i \g/2 ] }}
 {{\sinh[v_m + \omega_k+i\g/2]}\o{\sinh[v_m +\omega_k-i \g/2 ] }} = \cr
&\prod_{k=1, k \neq m}^r
{{\sinh[v_m - v_k + i\g]}\o{\sinh[v_m -v_k- i \g ] }}
{{\sinh[v_m + v_k + i\g]}\o{\sinh[v_m +v_k- i \g ] }} \cr
&\qquad 1 \leq m  \leq r}
\eqn\baes
$$
These Bethe Ansatz Equations (BAE) guarantee the vanishing
of the so-called unwanted terms. That is, vectors in
$t(\t,{\tilde \omega}) \Psi (\vec{v})$
which are not proportional to $\Psi (\vec{v})$ . They arise,
together with the wanted terms, from the repeated
commutation  of  $t(\t,{\tilde \omega})$  with
$ {\B}(v_1 ) {\B}(v_2 ) \ldots  \nxl {\B}(v_r )$
by means of the Yang-Baxter algebra \ybau\ .
Let us point out that the order of the factors ${\B}(v_j )$
in eq.\abal\  is irrelevant since the Yang-Baxter algebra \ybau\
implies
$$
\left[ \; {\B}(\t) ,\;  {\B}(\t') \right] = 0
\eqn\cobb
$$
{}From eq.\bsom\  it is evident that  $ {\B}(\t)$
is an odd function of  
$$
 {\B}(-\t) = -  {\B}(\t)
\eqn\bimp
$$
Consequently, a direct check shows that the BAE \baes\  are
invariant under negation of any single unknown  $v_j$ .
This implies that it is enough to consider BAE roots with
strictly positive real parts. In particular, purely imaginary pairs
($\pm i\eta$)  are ruled out. Moreover, one verifies by
 explicit calculation that  ${\B}(i\pi/2) \Omega =  0$ .
So that the selfconjugate root  v = i9/2  is also ruled out.

The eigenvalue of   $t(\t,{\tilde \omega})$   on  $\Psi (\vec{v})$
  is given by
$$
\Lambda (\t ; {\vec v}\, ) = \Lambda_+(\t ; {\vec v}\, )
+ \Lambda_-(\t ; {\vec v}\, )                    \eqn\eigev
$$
$$\eqalign{
\Lambda (\t ; {\vec v}\, ) =&\; {{\sinh(2\t \pm i \g)}\o{\sinh(2\t + i \g)}}
\; \prod_{k=1}^N
     {{\sinh[\t - \omega_k+i\g/2]}\o{\sinh[\t -\omega_k-i \g/2 ] }}
 {{\sinh[\t + \omega_k+i\g/2]}\o{\sinh[\t +\omega_k-i \g/2 ] }}  \cr
&\prod_{m=1}^r
{{\sinh[\t - v_m \mp i\g]}\o{\sinh[\t - v_m ] }}
{{\sinh[\t + v_m \mp  i\g]}\o{\sinh[\t + v_m ] }} \cr}
\eqn\avab
$$
Taking logarithms, eq.\avab\  become
$$\eqalign{
  & \sum_{k=1}^N \left[\phi(v_m - \omega_k,\g/2)+\phi(v_m +
\omega_k,\g/2 )\right] \cr  &=
  2 \pi I_m + \sum_{k=1,k\neq m}^r \left[\phi(v_m - v_k ,\g) +
 \phi(v_m + v_k ,\g) \right]\cr
 & \qquad 1 \leq m  \leq r}
\eqn\lbae
$$
where the  $I_m$  are positive integers[\chico ] and
$$
\phi(\l,x) \equiv i\log{{\sinh(ix+\l)}\o{\sinh(ix-\l)}}
                         \qquad  {\rm with}~~     \phi(0,x)=0.
  \eqn\defi
$$
Before solving these equations in the $N \to \infty$ limit,
let us study the $\t \to \infty$ behavior of
$\Lambda (\t ; {\vec v}\, )$ in order
to match with the asymptotics \ascal\ . We find from
eqs.\eigev -\avab
$$
\Lambda (\t ; {\vec v}\, )\buildrel{\t\to \infty}\over =
2 \cos\left[\g(N+1-2r)\right]~ [ 1 + O(e^{-2\t})]
\eqn\asil
$$
Here all roots  $v_1, v_2, ...., v_r$ are assumed to be finite.
Actually this must be the case unless $\g/\pi$  is
rational (see sec.5). Comparison of eqs.\asitu\  and \asil\
shows that there is only one positive (or zero) solution
$$
J = {N \o 2} - r
\eqn\jota
$$
except when  $\g/\pi$  is rational where other possibilities
may happen. In addition, we have from eqs.\ccaj\  and \refe\
that
$$
J_z \Psi (\vec{v}) = \left( {N \o 2} - r \right)\Psi (\vec{v})
\eqn\jotz
$$
Eqs.\jota\  and \jotz\
show that $J = J_z$ for Bethe-Ansatz eigenvectors. That is,
they are {\bf maximal weight vectors for the quantum group} for
non-rational values of  $\g/\pi$  . Thus,
$$
J_+ \Psi (\vec{v}) = 0
\eqn\jotm
$$
In addition, the kernel of  $J_+$  is stable under
variations of  $\g$ . Therefore eq.\jotm\  holds even when $\g/\pi$
is rational (see below). Each eigenvalue $\Lambda (\t ; {\vec v}\, )$
  has (at least) a degeneracy (2J+1) for   $t(\t,{\tilde \omega})$
, since the vectors
$$
J_- \Psi (\vec{v}),\; (J_-)^2 \Psi (\vec{v}),\ldots ,\;(J_-)^{2J} \Psi
(\vec{v}),
$$
are all eigenvectors of   $t(\t,{\tilde \omega})$ with the same eigenvalue
thanks to eqs.\tjme -\tjma.

Of course, when   $\g/\pi$ is rational  J  {\bf may not be equal} to  $J_z$
for
some BA states   $\Psi (\vec{v})$  as follows from eqs..\asitu ,\asil\
and \jotz .

Let us give an algebraic proof of the highest weight condition \jotm\
for the BA states. We shall apply  $J_+$  to $\Psi (\vec{v})$
given by eq.\abal\  permuting  $J_+$  through the   ${\B}(v_j
) , ( 1 \leq j \leq r)$
 with the help of eq.\bjma\ . Finally we shall use that
$$
J_+ \Omega = 0
\eqn\trivi
$$
Actually it is more convenient to use the operator ${\D}(\t)$
$$
{\D}(\t)= { 1 \o {\sinh(2\t - i \g)}}
\left[ {\cal D}(\t) \sinh 2\t -  {\cal A}(\t) \sinh(i\g) \right]
\eqn\dsom
$$
Eq.\bjma\  can now be written as
$$
 J_+ \; {\B}(\t) = e^{i\g} \; {\B}(\t) \; J_+ +
[e^{-\t}~ {\cal A}(\t)- e^{\t}~ {\D}(\t) ]
\eqn\jbso
$$
We find from eq.\abal\  after using r times eq.\jbso\ :
$$\eqalign{
J_+ \Psi (\vec{v})=& \sum_{j=1}^r \exp[i\g({{N+3}\o2} - r + 2j)]\;
 {\B}(v_1) {\B}(v_2)..... {\B}(v_{j-1}) \cr
&\left[ e^{-v_j} {\cal A}(v_j ) - e^{v_j } {\D}(v_j)\right]
 {\B}(v_{j+1}).... {\B}(v_r)\Omega \cr}
\eqn\gorda
$$
where we also used
eqs.\jotz\  and \trivi\ . Now we can push the operators $ {\cal A}(v_j )$
  and  $ {\D}(v_j)$  in eq.\gorda\  to the right using the Yang-Baxter
algebra \ybau . That is,
$$\eqalign{
 {\cal A}(\t)\; \B (\t') =& \; {{\sinh(\t-\t'-i\g) \sinh(\t+\t'-i\g)}
 \o {\sinh(\t-\t') \sinh(\t+\t')}} \; \B (\t')\; {\cal A}(\t)~+ \cr
  & {{\sinh(2 \t - i \g ) \sinh(i\g )}\o {\sinh(2\t )}} \; \B (\t)
\left[ {{{\cal A}(\t')}\o { \sinh(\t-\t')}} - { {\D (\t')} \o { \sinh(\t+\t')}}
\right] \cr
 {\D }(\t)\; \B (\t') =& \; {{\sinh(\t-\t'+i\g) \sinh(\t+\t'+i\g)}
 \o {\sinh(\t-\t') \sinh(\t+\t')}} \; \B (\t')\; {\D}(\t)~+ \cr
  & {{\sinh(2 \t + i \g ) \sinh(i\g )}\o {\sinh(2\t )}} \; \B (\t)
\left[ {{{\cal A}(\t')}\o { \sinh(\t+\t')}} - { {\D (\t')} \o { \sinh(\t-\t')}}
\right] \cr}
\eqn\abdb
$$
We find from eqs.\gorda -\abdb\  an expression of the form
$$
J_+ \Psi (\vec{v})=\sum_{j=1}^r c_j({\vec v})~
{\B}(v_1) {\B}(v_2)..... {\B}(v_{j-1}) {\B}(v_{j+1}).... {\B}(v_r)\Omega
\eqn\demo
$$
where the $ c_j({\vec v})$    are c-number coefficients.
 The easier is to first compute   $ c_1({\vec v})$  and then deduce
the rest by symmetry of the arguments (cfr. eq.\cobb\ ) .
  $ c_1({\vec v})~ $  follows by permuting in the first term of eq.\gorda\
the bracket $\left[ e^{-v_j} {\cal A}(v_j ) - e^{v_j } {\D}(v_j)\right]$
through $ {\B}(v_2)....  {\B}(v_r)$
using eqs.\abdb\  and keeping just
the first terms in eqs.\abdb\ . That is, those proportional
 to $ {\B}(v_j) \;  {\cal A}(v_1 )$ and to  $ {\B}(v_j) \; {\D}(v_1 ) ,
 \; (2 \leq j \leq r) $.
 Collecting all factors, it is easy to see that   $ c_1({\vec v})$
and the   $ c_j({\vec v}) \;  (1 \leq  j \leq  r)$ are proportional to the
 BAE \baes\  and hence identically zero. We have therefore
 proved that all BA vectors are highest weights [\ddvI ,\leon ,\maww ].

\chapter{The quantum group covariant ligth-cone approach.}

        The light-cone approach is a direct way to give
 a field theoretical interpretation to a lattice model
and furthermore obtain its scaling limit as a massive QFT.
We start from a diagonal lattice that we interpret as a
discretized two-dimensional Minkowski spacetime.
The sites in the light-cone lattice are considered as world events.
 Each site has then microscopic amplitudes associated to
it which describe the different processes that may take
place and must verify general properties like unitarity.

 		The light-cone approach with periodic
boundary conditions has been investigated in refs.[\ddv
-\ddvb ].
Here we consider this approach in the case of fixed boundary
conditions leading to a quantum group invariant framework.
For the six-vertex model, this can be done as follows.
Consider the  $SU(2)_q$  invariant transfer matrix  $t(\t,{\tilde \omega})$
constructed in sec.2, and identify the inhomogeneity
parameters  $ \omega_1,  \omega_2,....,  \omega_N$  with the cutoff rapidities
of the diagonal lattice. Namely set
$$
 \omega_j = (-1)^j~\Th
\eqn\inho
$$
Then define
$$
U(\Th) =\; t(\Th + i \g / 2 ,  \omega_j = (-1)^j~\Th )~
{{\sinh( 2 \Th + i \g )} \o {\sinh( 2 \Th + 2 i \g )}}
\eqn\defu
$$
We shall now show that  $U(\Th)$  has the natural interpretation
of the unit time evolution operator on the diagonal lattice.
To this end we introduce some convenient compact notation.
 We denote by  $0$  the auxiliary horizontal space and label
by $ j ,\; ( 1 \leq j \leq N)$,  the vertical spaces. Then
$$
L_j (\t) = R_{0j}(\t -\omega_j)~P_{0j}
\eqn\opel
$$
is a local vertex acting
only on the  jth.  vertical space. Here  $R_{jk}(\t)$  stand
for the six-vertex R-matrix \rma\  acting on the spaces
$j$  and  $k \; , ( 0 \leq j , k \leq N)$  and  $P_{jk}$  is the usual
exchange operator. In practice, the  $ab$  matrix element
 of  $L_j (\t)$  in the  0th. space act on the jth.  space as
 the operators  $t_{ab}(\t -\omega_j)$  defined in sec.2. In terms of
 the  $L_j (\t)$    we can write the transfer matrix as
$$
 t(\t + i \g / 2 , {\tilde \omega} ) = Tr_0 \{ K_0^+( \t + i \g)
L_N (\t).....L_1 (\t) K^-_0(\t) L_1 (-\t)^{-1}.....L_N (-\t)^{-1} \}
\eqn\prol
$$
Since the R-matrix is regular,$ R(0) = {\bf 1}$ , we have
$$
L_j (\t)\mid _{\t =\omega_j} =P_{0j}
\eqn\lome
$$
Hence, as soon as $ \omega_j = (-1)^j~\Th$ , one obtains
$$
L_{2k} (\Th) =P_{0,2k} \quad , \quad
L_{2k+1} (\Th) =S_{0,2k+1}
\eqn\lthe
$$
where  $S_{jk} = R_{jk}(2\Th) P_{jk}$ .
Inserting eq.\lthe\  into eq.\prol\  yields for odd N
$$\eqalign{
 & t(\t + i \g / 2 ,  \omega_j = (-1)^j~\Th )  \cr
=& Tr_0 \{ K_0^+( \t + i \g)\;
S_{0,N}P_{0,N-1}S_{0,N-2}P_{0,N-3} .....S_{0,3}P_{0,2}S_{0,1} K^-_0 \cr
& P_{0,1}S_{0,2}P_{0,3}......S_{0,N-1} P_{0,N} \}  \cr
=& Tr_0 \{ K_0^+( \t + i \g)\;
S_{0,N}S_{N-2,N-1} .....S_{1,2}P_{0,N-1}P_{0,N-3}......P_{0,2}K^-_0(\Th)\cr
& P_{0,1}P_{0,3}......P_{0,N}S_{2,3}......S_{N-1,N} \} \cr
=& Tr_0 \{ K_0^+( \t + i \g) \; S_{0,N}P_{0,N} \}\;
S_{12}.....S_{N-2,N-1} P_{0,N}P_{0,N-1}.....P_{0,2} \cr
& P_{0,1}P_{0,3}....P_{0,N} K^-_1(\Th)  S_{23}.....S_{N-1,N} \cr
=& Tr_0 \{ K_0^+( \t + i \g) \; R_{0,N}(2\Th)\}\; R_{12}(2\Th).....
R_{N-2,N-1}(2\Th)\cr & K^-_1(\Th)\; R_{23}(2\Th).....
R_{N-1,N}(2\Th) \cr}
\eqn\bodr
$$
Actually, the computation up to now is
completely general and would hold for any R-matrix and
 K-matrices. Let us exploit now the explicit form of
$R(\t)$ and $K^{\pm}(\t)$  for the six-vertex model
[eqs.\rma -\soldi ]. Specialising to the $SU(2)_q$
invariant case, we obtain
$$
K^-_1(\Th) = g_1(\Th)^{-1} \quad , \quad
Tr_0 \{ K_0^+( \t + i \g) R_{0,N}(2\Th)\} = g_N(\Th) ~
{{\sinh( 2 \Th + i \g )} \o {\sinh( 2 \Th + 2 i \g )}}
\eqn\kuno
$$
where
$$
g(\t) = \exp[-\t \s_z] =\pmatrix{e^{-\t}&0\cr
                    0&e^{+\t}\cr}
\eqn\mage
$$
Thus
$$
   U(\Th)=R_{12}R_{34}\ldots R_{\Ns-1-\e,\Ns-\e} \sub gN g_1^{-1}
                 R_{23}R_{45}\ldots R_{N-2+\e,\Ns-1+\e}          \eqn\uop
$$
where $\e=[1-(-1)^N]/2$, $\; g_j \; $ is the matrix $ \exp[\Th\s^z] $ acting
nontrivially only on the two-dimensional vector space attached to the $j^{th}$
link, and $R_{jk}$ is the 6V R--matrix $R(2\Th)$ acting nontrivially only on
the tensor product of the $j^{th}$ and $k^{th}$ vector spaces.

$ U(\Th) $  is clearly the unit time evolution operator on a diagonal
lattice with "reflecting"  type boundary conditions,
$g(\t)\;[g(\t)^{-1}]\;$ acting for collisions on the right [left]
 wall. Graphically  $ U(\Th) $  can be depicted as in fig.6.
 By taking powers of $ U(\Th) $ one then obtains the evolution
 on the diagonal lattice for any discrete time (Fig.7).
By construction, this time evolution is  $SU(2)_q$  invariant.

The eigenvectors of   $ U(\Th) $  can now be written as in sec.3,
namely
$$
\Psi (\vec{v}) =  {\B}(v_1 ) {\B}(v_2 ).... {\B}(v_r )
\Omega
\eqn\abald
$$
where   $ \omega_j = (-1)^j~\Th$
  and the Bethe Ansatz equations \baes\  take the form
$$\eqalign{
& \left[ {{\sinh[v_m - \Th +i\g/2]}\o{\sinh[v_m -\Th -i \g/2 ] }}
 {{\sinh[v_m + \Th +i\g/2]}\o{\sinh[v_m +\Th -i \g/2 ] }} \right]^N = \cr
&\prod_{k=1, k \neq m}^r
{{\sinh[v_m - v_k + i\g]}\o{\sinh[v_m -v_k- i \g ] }}
{{\sinh[v_m + v_k + i\g]}\o{\sinh[v_m +v_k- i \g ] }} \cr
&\qquad 1 \leq m  \leq r}
\eqn\baet
$$
The associated
eigenvalue of   $ U(\Th) $  is given by
$$
\U(\Th ,v_1,v_2,\ldots ,\sub vM) = \prod_{k=1}^r
{{\sinh[\Th- v_m - i\g/2]}\o{\sinh[\Th -v_m + i \g/2 ] }}
{{\sinh[\Th+v_m - i\g/2 ]}\o{\sinh[\Th + v_m + i \g/2 ] }}
\eqn\autu
$$
Let us now discuss the problem of unitarity for the time evolution
defined by   $ U(\Th) $ . The R--matrix \rma\ is unitary for real $\t$.
Hence the evolution operator $U$ is unitary too, for real $\Th$,
up to the boundary effects due to the term $\sub gN g_1^{-1}$ in eq.\uop.
As  is evident from figs.(6-7)
[eqs.\uop ] there exist unitarity violating
boundary effects (factors  $g(\t)_N$ and $g(\t)_1^{-1}$). Then,
the reflection of the bare particles on the boundaries
is affected by "leaking". However, on average this leaking
compesates since we have opposite factors  $g(\t)_N$ and  $g(\t)_1^{-1}$ on
right and left sides respectively. Thus one can expect   $ U(\Th) $
to still have unimodular eigenvalues. Indeed, this is the case
from eq.\autu\  provided, as it is usually the case, the BA roots
are either real, selfconjugate ($\Im v = \pi/2$) or organised in
complex conjugated pairs. As seen below,
the exact diagonalization of $U(\Th)$,
shows that all its eigenvalues are unimodular, so that there
exists a similarity transformation mapping $U(\Th)$ to an explicitly unitary
operator. On the other hand, it is natural to expect that in the thermodynamic
limit ($N\to\infty$) different boundary conditions become equivalent (the model
has a finite mass gap). This suggest that the above mentioned similarity
transformation reduces to the identity as $N\to\infty$. We conclude therefore
that the light--cone 6V model with fixed b.c. described by the evolution
operator $U$ of eq.\uop\ is another good, integrability--preserving
regularization of the SG model. Its advantage over the more conventional setup
with periodic b.c. is that $U(\Th)$ is explicitly $SU(2)_q$ invariant even for
finite $N$ .

According to the general light-cone
construction the physical lattice hamiltonian can be defined in terms of
 $U(\Th)$  as
$$
H = {i \o a}\log U(\Th)
\eqn\hami
$$
where  a  is the lattice spacing. By a judicious choice of the
logarithmic branches the energy
eigenvalues can be written from eqs.\autu -\hami\
$$
             E=a^{-1}\sum_{j=1}  {\rm e}_0(v_j)
\eqn\ene
$$
where
$$
            {\rm e}_0(v)= -\pi+2\arctan\left(
          {{\cosh2\Th\cos\g-\cosh2v}\o{\sinh2\Th \; \sin\g}}\right)
\eqn\ezero
$$
Notice that ${\rm e}_0(v)$ is smooth and negative definite for real $v$. The
specific choice of the logarithmic branch in passing from eq.\autu\
to eqs.\ene,  \ezero\ is dictated by the requirement that ${\rm
e_0}(v)$ should
correspond to the  negative energy branch of the spectrum of a single spin wave
over the ferromagnetic state $\Omega $.

The physical ground state and the particle--like excitations are obtained by
filling the interacting sea of negative energy states. This sea is described
by a set of real $v-$roots of the BAE with no ``holes" (in the Appendix we give
a detailed exposition on the treatment of the BAE, also to clarify some rather
subtle matters). Excitations correspond to solutions of the
BAE which necessarily contain holes and possibly complex
$v-$roots. The crucial point is that in the limit $N\to\infty$ only the number
of real roots of the sea grows like $N$, while the number of holes and complex
roots stays finite to guarantee a finite energy above the ground state
energy. Hence these solutions of the BAE can be described by densities
$\rho(v)$ of real roots plus a finite number of parameter associated to the
positions $x_1,x_2,\ldots x_\nu$ of the $\nu$ holes and to the location of
the complex $v-$roots.

The continuous theory is defined by the double limit  $a \to 0,
\Th \to \infty $  taken in such a way that  a finite mass gap  m  emerges.
The explicit result in the periodic case which will be shown
to hold also here is [\rev,\ddv ,\ddvb ]
$$
m = \lim_{a \to 0~\Th \to \infty} \left[ {4 \o a} e^{-{\pi \o a}\Th } \right]
\eqn\masa
$$
[The double limit is taken such that the l.h.s. is finite and non-zero].
The invariance of  $U(\Th)$  under the quantum group  $SU(2)_q$  allows
 to perform a transformation of the original vertex model where
states are assigned to the links and interactions as 'bare'
scattering amplitudes, to the vertices, into a diagonal lattice face model,
where states are defined on the plaquettes. The plaquettes
are the sites of the dual lattice and the interaction involves
the four plaquettes around each vertex, or equivalentely, the
four sites around each face of the dual lattice.

 On the faces of the diagonal
lattice one introduces positive integer--valued local height variables
$\ell_n(t)$, $t\in \ZZ$, according to (see fig. 2)
$$\eqalign{
           \ell_0(t)&=1\,;\;\ell_1(t)=2   \cr
           \ell_{n+1}(t)&=\ell_n(t)\pm 1\;,\; n=1,2,\ldots N-1\cr}   \eqn\locl
$$
The configurations $\{\ell_1(t),\ldots \ell_\Ns(t)\}$ at any fixed discrete
time
$t$ are in one--to--one correspondence with the $SU(2)_q$ multiplets of the 6V
model of dimension $\ell_n(t)=2J+1$ (notice that $\ell_\Ns(t)$ can be chosen to
be time--independent thanks to $SU(2)_q$ invariance). Now the matrix elements
of $U$ between these highest weight states define the unit--time evolution in
the face language.

This vertex-IRF correspondence is well known : in practice, it amounts
 to an application of the q-analog of the Wigner-Eckart theorem.
 Indeed, by  $SU(2)_q$  invariance, the matrix elements of   $U(\Th)$
in the subspace with definite total q-spin  $J = j$ , for  $\g/\pi$
 not a rational, can be written
$$
<j m ; \tau | U(\Th) | j' m' ; \tau' > = \delta_{jj'}\;\delta_{mm'}~
<\tau|| U(\Th) ||\tau'>_j
\eqn\tweq
$$
where  $m = -j, -j + 1, ...., +j$  labels the  $J_z$  eigenvalues and
 $\tau$  the degeneracy of the  $J = j$  subspace which has dimension
$$
\pmatrix{N \cr N/2 - j \cr} - \pmatrix{N \cr N/2 - j -1  \cr}
\equiv \Gamma^{(N)}_j
\eqn\dime
$$
$<\tau|| U(\Th) ||\tau'>_j$ thus denotes the reduced matrix elements,
 and a specific choice of basis is required to give explicit
expressions for them. The most natural basis, which is the
basis useful for the vertex-IRF correspondence, is that obtained
by the successive composition of the q-spin 1/2 basic constituents
assigned to the links. Consider a 'time zero' line cutting the
 diagonal lattice as in fig.8. By intersecting N links, it
identifies the Hilbert space of the vertex model as $ (\otimes [1/2] )^N$
, where $ [j] $  denotes an irreducible representation of weight  j .
On the other hand, the line passes also through a well defined set
of plaquettes, including the half-plaquette at the extreme
left and right. Then, a set of configurations of a local
height variable  {\it l} , assigned to each plaquette, can be
constructed as follows. Assign   ${\it l}  =   {\it l}_0  = 1$  to the half
plaquette on the left, the 0th. plaquette. Then pass to the
next one on the right , the first plaquette, following the time
zero line. This cuts a link carrying a spin 1/2 representation.
So that, at this stage the Hilbert space is just the
 representation  $[1/2]$  itself, i.e. $ J = 1/2$ . We then set
  ${\it l}_1 = 2 J + 1 = 2$ . Following the line, we now cut another
spin 1/2 link and arrive at the second plaquette.
This Hilbert space is now the direct sum $ [J=0] \oplus [J=1]$
and the local height l can take two values,  $ {\it l}_2 = 2j + 1 = 1$ or 3 .
 By repeating this procedure  n  times (with  $n \leq N$),
we land in the  nth. plaquette after having cut $n$  spin 1/2 links :
the Hilbert space is    $ (\otimes [1/2] )^n$  and contains irreps with  J
running from  0  or  1/2  up to $ n/2$ . Hence the local variables ${\it l}_n$
  take the values
$$
 {\it l}_n = 2 J + 1 = 	\left\{
\matrix {1, 3, 5,\ldots, n + 1 ,~~  n {~\rm even} \cr
        2, 4, 6,...... , n + 1 ,~~  n {~\rm odd } }\right.
$$
By construction,  $ |{\it l}_n - {\it l}_{n-1}| = 1 $ ,
 since the  ${\it l}_n$  follow the composition laws
 of  $SU(2)_q $ representations
$$
[J] \otimes  [1/2]  = [ J + 1/2] \oplus  [J - 1/2]
$$
which for  $q = \exp(i\g)$  not a
root of unity, are just those of the usual SU(2).

When  $n = N$  we arrive at the half-plaquette on the right.
 Here the values of  ${\it l}_n$  are just the dimensions of the
irreducible representations in which the full Hilbert space  $(\otimes [1/2]
)^N$
  can be decomposed. All together, a
specific choice of  $ {\it l}_2,  {\it l}_3,\ldots,{\it l}_N $ (
 $ {\it l }_0 $ and   ${\it l}_1 $
are fixed to   $ {\it l }_0 = 1 ,~  {\it l }_1 = 2 $ by construction) defines
one and only one of the  $\Gamma_j(N)$ irreps with  $J = j$. , if
 ${\it l}_N = 2j + 1$ . In other words, we have constructed a map
from the  N + 1  plaquettes at 'time zero' and the Bratteli
 diagrams giving the composition rules for the tensor
product of  N  spin 1/2 reprersentations (see fig.9).
We are now in position to identify the degeneracy label  $\tau$
with a specific path in the Bratelli diagram, i.e.
 $\tau = ( {\it l}_0,  {\it l}_1,\ldots,l_N )$ , with $ {\it l }_0 = 1$
 and  $l_N  = 2j + 1$ .

It remains to evaluate the reduced matrix elements
$$
<\tau|| U(\Th) ||\tau'>_j\;  =\;  < {\it l}_0,  {\it l}_1,\ldots,{\it l}_N|
 U(\Th) | {\it l'}_0,  {\it l'}_1,\ldots,{\it l}_N>
\eqn\emre
$$
where the height variables $ {\it l'}_0,  {\it l'}_1,\ldots,{\it l}_N$
are associated to the plaquettes crossed by the 'time one'
line (see fig.8). Notice that $  {\it l}_0 =   {\it l'}_0 =  1$  by definition,
while  $ {\it l}_N =   {\it l'}_N  = 2j + 1 $ by  $SU(2)_q$  invariance.
However, in spite of the factorised form of  $ U(\Th)$
[cft. eqs.\uop ], these matrix elements cannot be
written in a factorised form, with each factor depending
locally on the variables $ {\it l}_0,  {\it l}_1,\ldots{\it l}_N$
 and  $ {\it l'}_0,  {\it l'}_1,\ldots,{\it l}_N $ .
This is so because the full unit time evolution operator   $ U(\Th)$
is $SU(2)_q$  invariant, but each single  $R_{k,k+1}(2\Th)$ which enters is
not.
Only the peculiar boundary terms  $g_1$  and  $g_N$  enforce  $SU(2)_q$
invariance. Nevertheless, it is easy to pass to a different vertex
representation where the $SU(2)_q$  symmetry holds in a local sense.
We define a similarity transformation on   $ U(\Th)$
$$
{\tilde U}(\Th)= G U(\Th)G^{-1}
\eqn\gugm
$$
where
$$
G= g_1^{1/2}~ g_2^{-1/2}~ g_3^{1/2}\ldots g_N^{\epsilon -1/2}
\eqn\gexp
$$
$\epsilon = {{1 - (-1)^N}\o 2}$  and  $g_j$  is $\exp[-\Th \s_z]$
acting on the j-th.
spin 1/2 space. Now comparing eqs.\uop\  and \gugm -\gexp\
it is easy to verify that  ${\tilde U}(\Th)$  can be written, e. g.
for even N
$$
  {\tilde U} (\Th)={\tilde R}_{12}{\tilde R}_{34}\ldots
{\tilde R}_{N-1,N} {\tilde R}_{23}{\tilde R}_{45}\ldots
{\tilde R}_{N-2,N-1}          \eqn\uopt
$$
where
$$
{\tilde R}_{k,k+1}=g_k^{-1/2}g_{k+1}^{1/2} R_{k,k+1}(2\Th)
g_k^{1/2}g_{k+1}^{-1/2}
\eqn\rtil
$$
One can explicitly verify that  ${\tilde R}_{k,k+1}$  is
$SU(2)_q$  invariant; that is
$$\eqalign{
[{\tilde R}_{k,k+1} ,\;  (\s_z)_m\;  + \;  (\s_z)_{m+1} ] =& 0~~~, \cr
[{\tilde R}_{k,k+1} ,\; (q^{-\s_z/2})_m \; (\s_{\pm})_{m+1}\;  +\;
(\s_{\pm})_{m}
\; (q^{\s_z/2}) ]
=& 0 \cr }
\eqn\invq
$$
The reduced matrix elements of $ {\tilde U} (\Th)$
can now be expressed in a factorised form ,
$$
 < {\it l}_0,  {\it l}_1,\ldots,{\it l}_N|
 U(\Th) | {\it l'}_0,  {\it l'}_1,\ldots,{\it l}_N>\;  =\;  W'_1 W'_3  W'_5
\ldots  W'_{N-1+\epsilon} W_2 W_4 \ldots  W_{N-\epsilon} ~,
$$
where
$$
W_m = W({\it l}_{m-1},{\it l}_{m+1}|{\it l}_{m},{\it l'}_{m};\Th)~~,~~
W'_m = W({\it l'}_{m-1},{\it l'}_{m+1}|{\it l}_{m},{\it l'}_{m};\Th)
\eqn\dobw
$$
and
$$
W(x,y|u,v;\Th)=\delta_{uv} - {{\sinh2\Th}\o{\sinh(2\Th+i\g)}}
{{\sqrt{[u]_q[v]_q}}\o{[x]_q}}~\delta_{xy}.
\eqn\peso
$$
with the standard notation
$ [x]_q  =  (q^x - q^{-x}  ) / ( q - q^{-1} )  = \sin( \g x ) / \sin \g $.
This completes the transformation from vertex to faces
description. The global time evolution in the faces
(or heights) language is obtained by taking matrix products of
$$
 < {\it l}_0,  {\it l}_1,\ldots,{\it l}_N |
 U(\Th) | {\it l'}_0,  {\it l'}_1,\ldots,{\it l}_N>
$$
The boundary conditions are  $ {\it l}_0 = {\rm constant}  =  1$  on the left,
while  ${\it l}_N  = {\rm constant} = 2j + 1$ on the right, if the reduction
is performed onto the  $\Gamma_j^{(N)}$- dimensional space of irreps of
q-spin  J = j . With the natural constraint that
$|{\it l - l}'|  = 1$ , when  {\it  l}  and   {\it l}'
sit on neighboring faces,
the weights given by eqs.\dobw -\peso\  define the ABF-SOS model
in the trigonometric regime[\abf ].

\chapter{Analysis of the Bethe Ansatz equations}

	We investigate in this section the solution of the BAE
\baet\  associated to
the quantum group covariant BA in the light-cone approach.
It is convenient to relate them with the BAE for periodic
boundary conditions (see for example [\rev ]).
	Define  2r  variables  $\l_j$  as
$$
\l_j = v_j \qquad, \qquad \l_{j+r} = -v_{r-j+1}~,\qquad 1 \leq j \leq r
\eqn\lamd
$$
Then eqs.\baet\
can be rewritten as
$$\eqalign{
& \left[ {{\sinh[\l_m - \Th +i\g/2]}\o{\sinh[\l_m -\Th -i \g/2 ] }}
 {{\sinh[\l_m + \Th +i\g/2]}\o{\sinh[\l_m +\Th -i \g/2 ] }}\right]^N
{{\sinh(2 \l_m + i \g)}\o{\sinh(2 \l_m - i \g)}}
= \cr
&-\prod_{k=1}^{2r}
{{\sinh[\l_m - \l_k + i\g]}\o{\sinh[\l_m -\l_k- i \g ] }}
{}~, \qquad   1 \leq m  \leq 2r .}
\eqn\baec
$$
These equations are like the BAE for periodic boundary conditions
on a  2N  sites line and with an additional source factor
$$
{{\sinh(2 \l_m + i \g)}\o{\sinh(2 \l_m - i \g)}}=
-{{\sinh(\l_m + i \g/2)}\o{\sinh(\l_m - i \g/2)}}
{{\sinh(\l_m - i {{\pi -\g}\o 2})}\o{\sinh( \l_m + i {{\pi -\g}\o
2})}}
\eqn\sinc
$$
More important, we have the following constraints on the
roots  $\l_m$ :

\item{a)} the total number of roots is even (2r) and they
are symmetrically distributed with respect to the origin according
to eq.\lamd .

\item{b)} $\l_m  = 0$  and  purely imaginary   $\l_m$  are excluded as
roots.

Let us start by considering the antiferroelectric
ground state. It is formed by roots with fixed imaginary part equal
to $\pi/2$ . Therefore, we shift  $\l_m  \to \l_m   + i\pi/2 $ and take
 $\l_m$ real for the ground state. Now  $\l_m = 0$ is excluded whereas it
may be present for periodic boundary conditions (PBC). Therefore,
the ground state for eq.\baet\ is a one-hole solution of eq.\baec\
with the hole at  $\l_m = 0$.We also redefine $\g$ into $\pi - \g$
in order to agree with the conventions of refs.[\rev ,\ddv ,\ddvb ].
With this choice, $\g$ is related to the sine-Gordon coupling constant
$\b$ by $\b ^2 /(8\pi) = 1 - \g/\pi $.

Let us now consider the  $N \to \infty$
limit where a continuous density of BAE real roots can be introduced
$$
\rho(\l_m ) = \lim_{N \to \infty} {1 \o{N(\l_{m+1} - \l_m )}}
\eqn\dens
$$
This function must always be even in
 our fixed boundary conditions case. Taking logarithms in both sides
of eq.\baec\  and using eqs.\sinc\  and \dens\  the BAE yield linear
integral equations through the usual procedure
$$\eqalign{
&N\left[ \Phi'(\l + \Th, \g/2) + \Phi'(\l - \Th, \g/2)\right]~
+ ~ \Phi'(\l , \g/2)~ - ~  \Phi'(\l , {{\pi -\g}\o 2}) \cr
&= ~ 2 \pi N \rho(\l) ~ + N \int_{-\infty}^{+\infty} d\mu \; \rho(\mu)
\; \Phi'(\l - \mu, \g) ~ + \cr
&\sum_{l=1}^{2N_c}\left[\Phi'(\l - z_l, \g)+\Phi'(\l - {\bar z}_l, \g)
\right] + 2\pi \delta(\l) + 2\pi \sum_{h=1}^{2M_h}\delta(\l - \t_h)\cr}
\eqn\cont
$$
where we assume  $2N_c$  pairs of complex roots  ($z_k,{\bar z}_k$)
 and $2M_h$  holes  $\t_h$ , symetrically distributed with respect to the
origin.
	The roots density  $\rho (\l)$  is connected with  $\s(\l)$ ,
 the derivative of the counting function through
$$
\rho (\l)=\s(\l) -{1 \o N}\delta(\l) -
{1 \o N}\sum_{h=1}^{2M_h}\delta(\l - \t_h)
\eqn\rysi
$$
The Fourier transform solution of eq.\cont\  gives for the
ground state
$$\eqalign{
\s_0(\l) &= \int_{-\infty}^{+\infty}{{dk} \o {2\pi}} ~e^{ik\l}
\;{\tilde \s}_0(k)\cr
{\tilde \s}_0(k) &= {{\cos(k\Th )+ 1/(2N)} \o {\cosh(k\g /2)}} +
{1 \o {2N}} {{\sinh[k(\pi - 3 \g)/4]} \o
{ \cosh(k\g /2) \sinh[k(\pi - \g)/4]}}\cr}
\eqn\denv
$$
Notice that the first term is twice the PBC ground state density
(cft. ref.[\rev ]).
The second term is the boundary effect and leads to a  $N^{-1}$
surface correction to the free energy[\aust ].

	The densities of real roots for excited states,
that is in the presence of (non-zero) holes and complex roots
are analogous to the PBC case.
We find from eq.\cont\ :
$$
\s(\l) = \s_0(\l) + {1\o N}[\s_h(\l) + \s_c(\l) ]
\eqn\densi
$$
Here $\s_h(\l)$  and  $\s_c(\l)$  stand for the holes and complex
roots contributions, respectively. The explicit expressions of
these densities are:
$$
\s_h(\l) = \sum_{h=1}^{2M_h}p(\l - \t_h)\quad , \quad
\s_c(\l) = \sum_{h=1}^{2N_c}Q(\l - \s_l , \eta_l )
\eqn\denac
$$
where  $z_k  = \s_k + i \eta _k$  with  $ \eta _k > 0$  and
$$\eqalign{
 p(\l)=& \int_{-\infty}^{+\infty}{{dk}\o{2\pi}}\;{{\sinh(\pi/2-\g)k}\o
              {2\sinh(\pi-\g )k/2 \,\cosh(\g k/2)}} \;e^{ik\l} \cr
Q(\l , \eta) =& -p(\l - i \eta)-p(\l + i \eta)~ ,
{}~~ \eta < \g < \pi/2 .\cr
Q(\l , \eta) =&-  {1 \o {2\pi}}{d \o {d\l }}\left[
\Phi_{\g}(\l ,\eta-\g)-\Phi_{\g}(\l ,\eta) \right]~,~~~\pi/2>\g<\eta.\cr}
\eqn\pyql
$$
Here
$$
\Phi_{\g}(\l ,\eta) \equiv \Phi \left({{\l} \o {1 - {{\g} \o {\pi}}}},
{{\eta} \o {1 - {{\g} \o {\pi}}}} \right)
\eqn\figl
$$
Complex roots with  $\Im (z) > \g$  are called wide pairs
whereas roots in the region  $\Im (z) < \g$  appear as two-strings
($\Im z = \pm \g/2$) or in quartets[\nons ].

	Let us now derive the higher-level BAE.
We follow a procedure analogous to ref.[\nons ]
for the PBC case. The only novelty with respect to the PBC case
is the extra term $ -\Phi'( \l,(\pi-\g)/2)$  in the source and the
 obligatory presence of a hole at  $\l = 0$ . In order to derive
the higher-level BAE one has to compute
$$
     \s_0(\l) + \s_0(\l-i\g)
$$
We find from eq.\denv\
$$
 \s_0(\l) + \s_0(\l-i\g) = {1 \o {2 \pi N}}{d \o {d\l }}
\Phi_{\g}(2\l - i \g , \g)
\eqn\saig
$$
The r.h.s. comes entirely from the second term in eq.\denv\ .
 Notice that the l.h.s. of eq.\saig\  exactly vanishes in the
PBC case.

	We define the parameters  $\chi_j$  as in the PBC
 case[\nons ] :

$\chi$  = real part of the two-string position,

$\chi =  z_c - i \g/2, {\bar z}_c + i \g/2 $ for a quartet
$(z_c ,{\bar z}_c , z_c - i \g,{\bar z}_c + i \g )$,

$\chi  = z_w - i \g/2 , {\bar z}_w + i \g/2 $
for a wide pair $(z_w,{\bar z}_w)$.

Then, the BAE take the
following form in the  $N = \infty$  limit
$$\eqalign{
& \prod_{h=1}^{\nu} {{\sinh\a[\chi_j - \t_h +i\g/2]}
\o{\sinh\a[\chi_j - \t_h  -i \g/2 ] }} = \cr
&-e^{i\Phi(2\chi_j,\g)} \prod_{k=1}^{M_p}~
{{\sinh\a[\chi_j - \chi_k + i\g]}\o{\sinh\a[\chi_j - \chi_k- i \g ] }}
{}~, \qquad   1 \leq j  \leq M_p .\cr}
\eqn\nsan
$$
where  $\a \equiv  1/(1 - \g/\pi )$  and the
use of eq.\saig\  yields the phase factor
$e^{i\Phi(2\chi_j,\g)} $
  in the r.h.s. . The  $\nu  = 2M_h$  holes and the $M_p$  complex
pairs are all distributed symmetrically with respect to the
 $\Re\l = 0$  axis. Hence, restricting to positive real parts,
we have
$$\eqalign{
& \prod_{h=1}^{\M_h} {{\sinh\a[\chi_j - \t_h +i\g/2]}
\o{\sinh\a[\chi_j - \t_h  -i \g/2 ] }}
{{\sinh\a[\chi_j + \t_h +i\g/2]}
\o{\sinh\a[\chi_j + \t_h  -i \g/2 ] }}  = \cr
& \prod_{k=1,k\neq j}^{M_p/2}~
{{\sinh\a[\chi_j - \chi_k + i\g]}\o{\sinh\a[\chi_j - \chi_k- i \g ] }}
{{\sinh\a[\chi_j + \chi_k + i\g]}\o{\sinh\a[\chi_j + \chi_k- i \g ] }} \cr
& ~~ \qquad   1 \leq j  \leq M_p /2  .\cr }
\eqn\eufa
$$
To conclude, let us come back to the BAE \baet .
 They can be rewritten in a manifestly algebraic form
by introducing the variables  $z_j \equiv  \exp(-2v_j )$ ,
$(1 \leq j \leq r)$.
$$
     \left({{z_j w-q}\o{z_j wq-1}}\;{{z_j-wq}\o{z_j q-w}}\right)^N =
     \prod_{\scriptstyle k=1\atop\scriptstyle k\ne j}
     {{z_j-z_k q^2}\o{z_j q^2-z_k}}\;{{z_j z_k-q^2}\o{z_j z_k q^2-1}}
\eqn\baez
$$
where  $w = exp(-2\Th)$  and  $q = \exp(i\g)$  is the quantum group
deformation parameter. According to the general discussion
following eq.\bimp , we can restrict the search for solutions
to eq.\baez\  strictly within the unit circle  $|z_j | < 1$ .
Therefore, to any unordered set of distinct numbers fulfilling
eq.\baez ,  $z_1, \ldots , z_r$ , with   $|z_j | < 1$ ,
there corresponds one BA eigenstate of the transfer
matrix \prol\  and hence of the evolution operator on the
light cone-lattice. Special attention should be paid to
the possibility that one or more roots  $z_j$  lay exactly
at the origin. This corresponds to $\Re v_j \to \infty $ ,
and therefore to the reduction of the corresponding  $ {\B}(v_j )$
 to a multiple of the lowering operator $ J_-$  of $SU(2)_q$
[cft. eq.\ascal ].
$$
 {\B}(\infty) = (1 -q^2 ) J_-
\eqn\bj
$$
In effect, the point at infinity
constitutes the only exception to the requirement that
the roots  $v_1, \ldots , v_r$  be all distinct. However, this is
a very special possibility. Indeed, suppose that a given root
in eq.\baez\ , which we can always identify with $z_1$ , lays at
the origin  $z_1 = 0$ . We immediatly obtain from eq.\baez\ :
$$
q^{2N} = q^{4(r-1)} \Longrightarrow q^{4(J_z+1)} = 1
\eqn\qrau
$$
that is, $q$ must be a root of
unit and  $\g$  a rational multiple of $\pi$ . This correspond to
the special cases when certain irreducible representations of
$SU(2)_q$  mix into type I reducible but indecomposable
representations. This interesting
phenomena is discussed in secs.7-9 within the Bethe Ansatz
framework in connection with the RSOS models.

Eqs.\eufa\ can also be written in algebraic form:
$$
     \prod_{h=1}^\nu {{\xi_j w_h-{\hat q}}\o{\xi_j w_h {\hat q}-1}}\;
                {{\xi_j-w_h {\hat q}}\o{\xi_j {\hat q}-w_h}} =
     \prod_{\scriptstyle k=1\atop\scriptstyle k\ne j}^{\hat M}
     {{\xi_j-\xi_k {\hat q}^2}\o{\xi_j {\hat q}^2-\xi_k}}\;
     {{\xi_j \xi_k-{\hat q}^2}\o{\xi_j \xi_k {\hat q}^2-1}}     \eqn\hbae
$$
where
$$
      \xi_j=e^{-2\a\chi_j} \;,\;\; w_h=e^{-2\a x_h} \;,\;\;
      {\hat q}\equiv e^{i{\hat \g}}=e^{i\a\g}\;,\;\;
                                 \a={\pi\o{\pi-\g}}           \eqn\renpar
$$
These new BAE involve only a finite number of parameters and are formally
identical to the ``bare" BAE \baet -\baez, with the holes acting as sources in
the
place of the alternating rapidities $\pm\Th$. Moreover, from \hbae\ one reads
out the renormalization of the quantum group deformation parameter
$$
  q \to {\hat q} \quad  \ie \quad \g\to {\hat \g}={{\pi\g}\o{\pi-\g}} \eqn\ren
$$
As we shall see later on, this renormalization has a nice physical
interpretation for the SOS and RSOS models related to the 6V model by the
vertex--face correspondence.

\chapter{Higher--level Bethe Ansatz and S--matrix}

The holes in the sea of real BAE roots are the particles of the light--cone 6V
model. They are $SU(2)_q$ doublets and can be present in even (odd) number for
$N$ even (odd). Consider first the even $N$ sector, setting $N=2N^\prime$. The
energy of a BA state with an even number $\nu$ of holes located at
$x_1,x_2,\ldots x_\nu$ can be calculated to be, in the
$N^\prime \to\infty$ limit,
$$
    E=E_0+a^{-1}\sum_{h=1}^\nu {\rm e}(x_h)+O(a^{-1}N^{-1})\;,\quad
{\rm e}(x)=2\arctan\left({\cosh\pi x/\g}\o{\sinh\pi\Th/\g}\right)  \eqn\enhole
$$
where $E_0$ is the energy of the ground state, that is the state with no holes.
$E_0$ is of order $N$ and is explicitly given in the appendix along with some
detail on the derivation of eq.\enhole. Now suppose $N$ odd, with
$N=2N^\prime-1$. To compare this situation with the previous one, we need to
slightly dilate the lattice spacing $a$ to
$a^\prime=2N^\prime a/(2N^\prime-1)$, in order to keep constant the physical
size $L=Na$ of the system. Then eq.\enhole\ remains perfectly valid,
as $N^\prime \to\infty$, also for the case of $N$ odd, with the same ground
state energy $E_0$. The only difference is
that now $\nu$ is odd. Hence, alltogether, we obtain that the number of holes
can be arbitrary (unlike in the treatment with periodic b.c.) and that the
total energy, relative to the ground state and in the $L\to\infty$ limit, is
the uncorrelated sum of the energy of each single hole,
independently of the complex pair structure of the corresponding BAE solution.
The BA state with one hole has $J=N/2-M=(2N^\prime-1)/2-(N^\prime-1)=1/2$ and
is
therefore a $SU(2)_q$ doublet. The $2^\nu$ polarizations of a state with $\nu$
holes are obtained by considering all solutions of the higher--level BAE \hbae\
with $0\le{\hat M}\le \nu/2$. We see in this way that the holes can be
consistently interpreted as particles.

As a lattice system the light--cone 6V model is not critical. The
(dimensionless) mass gap is  the minimum value of the positive definite ${\rm
e}(x)$, the energy of a single hole, that is
$$
  {\rm e}(0)=2\arctan\left(1\o{\sinh\pi\Th/\g}\right)           \eqn\massgap
$$
This gap vanishes in the limit $\Th\to\infty$. Hence the continuum limit
$a\to0$ can be reached provided at the same time $\Th\to\infty$ in such a way
that the physical mass
$$
              m=a^{-1}{\rm e}(0) \simeq 4a^{-1}e^{-\pi\Th/\g}      \eqn\mass
$$
stays constant [\ddv]. In the same limit we obtain the relativistic expression
$$
            a^{-1}{\rm e}(x) \to m\cosh\pi x/\g                  \eqn\rel
$$
so that $\pi x/\g$ is naturally interpreted as the physical rapidity $\t$ of
the  hole. This is consistent also in our fixed b.c. framework, provided the
limit $L\to\infty$ is taken before the continuum limit. Indeed the total
momentum becomes a conserved quantity in the infinite volume limit and its
eigenvalues can be expressed in terms of the BA $v-$roots exactly as in the
periodic b.c. formulation. Then one finds that the momentum of a single
particle takes the required form $m\sinh\t$ in the continuum limit.

The above analysis shows that a relativistic particle spectrum appears in the
$a\to 0$ limit above the antiferromagnetic ground state. For $\g>\pi/2$ one
can show also that bound states appear, associated to appropriate strings of
complex roots, just as in the periodic b.c setup. In the sequel we shall anyway
restrict our analysis to the repulsive $\g<\pi/2$ region, where the only
particles are the holes, to be identified with the solitons of the SG model. Of
course at this moment, since the infinite volume limit is already implicit, the
particles are in their asymptotic, free states: the rapidities $\t_h=\pi
x_h/\g$ may assume arbitrary continuus values and the total excitation energy
is the sum of each particle energy and does not depend on the internal state of
the particles. The situation changes if we consider $L$ very large but finite,
since in this case the hole parameters $x_1,\ldots,x_\nu$ are still quantized
through the ``bare" BAE \baet.
Indeed, by definition the holes are real distinct
numbers satisfying
$$
          Z_N(x_h\,;v_1,v_2,\ldots,\sub vM)={2\pi{\bar I}_h \o N} \quad ;
                       \quad h=1,\ldots,\nu                    \eqn\quant
$$
where $Z_N(x\,;v_1,\ldots,\sub vM)$ is the ``counting function" defined in the
Appendix and the positive integers ${\bar I}_1,\ldots,{\bar I}_\nu$ are all
distinct from the integers $I_1,\ldots,I_r$ labelling the $r$ ($r\le M$) real
$v-$roots of the BAE. For $N$ very large, $J=N/2-M$ finite and $x<(\g/\pi)\ln
N$, the counting function can be approximated as
$$
    Z_N(x\,;v_1,\ldots,\sub vM)= Z_\infty(x)+
    N^{-1} F(x\,;x_1,\ldots,\sub xM;\sub{\chi}1,\ldots,\sub{\chi}M)
                         +O\left(N^{-2}\right)                 \eqn\zapp
$$
where $Z_\infty(x)$ is the ground state counting function {\it at} the
thermodynamic limit
$$
   Z_\infty(x)=2\arctan\left({\sinh\pi x/\g}\o{\cosh\pi\Th/\g}\right) \eqn\zinf
$$
and
$$\eqalign{
        F(x\,;x_1,\ldots,\sub xM; & \sub{\chi}1,\ldots,\sub{\chi}M)=
                              -i\log\prod_{h=1}^\nu
        S_0\big(\frac\g\pi(x-x_h)\big) S_0\big(\frac\g\pi(x+x_h)\big) \cr
                       &-i\log\prod_{j=1}^{\hat M}
         {{\sinh\a(x-\chi_j+i\g/2)\,\sinh\a(x+\chi_j+i\g/2)}
         \o{\sinh\a(x-\chi_j-i\g/2)\,\sinh\a(x+\chi_j-i\g/2)}}\cr}   \eqn\shift
$$
In the last expression, the numbers $\chi_j$ are the roots of the higher level
BAE \hbae, \renpar, while $S_0(\t)$ coincides with the soliton-soliton
scattering amplitude of the SG model
$$
     S_0(\t)=\exp\,i\int_0^\infty {dk\o k}
            {{\sinh(\pi/2\g-1)k}\o{\sinh(\pi k/2{\hat \g})}}
            {{\sin k\t/\pi}\o{\cosh k/2}}                          \eqn\solsol
$$
under the standard identification $\g/\pi=1-\beta^2/8\pi$.

Combining eqs. \quant\ and \zapp, and taking the continuum limit one obtains
the ``higher level" expression
$$
      \exp\left(-imL\sinh\t_h\right) =
                           \prod_{{n=1}\atop{n\ne h}}^\nu
        S_0(\t_h-\t_n) \;  S_0(\t_h+\t_n)    \; \prod_{j=1}^{\hat M}
      {{\xi_j w_h-{\hat q}}\o{\xi_j w_h {\hat q}-1}}\;
       {{\xi_j {\hat q}-w_h}\o{\xi_j-w_h {\hat q}}}         \eqn\renquant
$$
where, according to \renpar, $\xi_j=e^{-2\a\chi_j}$ and
$w_h=e^{-2{\hat \g}\t_h/\pi}$. Together with the higher level BAE \hbae, this
last equation provides the  exact Bethe ansatz diagonalization of the
commuting family formed by the $\nu$ renormalized one--soliton evolution
operators (see fig. 3)
$$\eqalign{
    {\hat U}_h=&S_{h,h-1}(\t_h-\t_{h-1})\ldots S_{h,1}(\t_h-\t_1)\,g_h(-\t_h)
               \,S_{h,1}(\t_h+\t_1)\ldots S_{h,\nu}(\t_h+\t_\nu) \times \cr
  &g_h(\t_h)\,S_{h,\nu}(\t_h-\t_\nu)\ldots S_{h,h+1}(\t_h-\t_{h+1})\cr}\eqn\one
$$
where $g_h(\t)=\exp\t\sigma_h^z$ and $S(\t)$ is the complete $4\times4$ SG
soliton $S-$matrix in the repulsive regime $\g<\pi/2$ (for brevity we set
${\hat \t}=\g\t/(\pi-\g)$):
$$
  \eqalign{S(\t)=S_0(\t)\pmatrix{1&0&0&0\cr
                                 0&{\hat b}&{\hat c}&0\cr
                                 0&{\hat c}&{\hat b}&0\cr
                                 0&0&0&1\cr} \cr}     \qquad
  \eqalign{{\hat b}=&\;b({\hat\t},{\hat\g})=
                 {{\sinh{\hat\t}}\o{\sinh(i{\hat\g}-{\hat\t})}} \cr
           {\hat c}=&\;c({\hat\t},{\hat\g})=
                 {{\sinh i{\hat\g}}\o{\sinh(i{\hat\g}-{\hat\t})}}\cr}  \eqn\sma
$$
The operators ${\hat U}_h$ are the values at
$\t=\t_h$ of the fully inhomogeneous Sklyanin--type transfer matrix
$T(\t\,;\t_1,\ldots,\t_\nu)$ constructed with $S(\t-\t_h)$ as local
vertices. Let us stress that the higher--level Bethe Ansatz structure just
described follows directly, after specification of the BAE solution
corresponding to the ground state and without any other assumptions, from the
``bare" BA structure of the light--cone 6V model. In particular, this provides
a derivation ``from first  principles" of the SG $S-$matrix. We want to remark
that $S(\t)$ is also the exact $S-$matrix for the elementary excitations of the
6V model {\it on the infinite lattice}: it is a bona--fide {\it lattice}
$S-$matrix. Of course, in this case $\t=\t_1-\t_2$ is the difference of
{\it lattice} rapidities, which are related to the energy and momentum of the
scattering particles through the {\it lattice} uniformization
$$\eqalign{
       P_j=&a^{-1}Z_\infty(\g\t_j/\pi)
        =2a^{-1}\arctan\left({\sinh\t_j}\o{\cosh\pi\Th/\g}\right) \cr
       E_j=&a^{-1}\e(\g\t_j/\pi)
        =2a^{-1}\arctan\left({\cosh\t_j}\o{\sinh\pi\Th/\g}\right) \cr}
\eqn\latu
$$
implying the {\it lattice} dispersion relation
$$
           \cos aE_j/2 =\tanh(\pi\Th/\g)\cos aP_j/2                    \eqn\lat
$$
Thus, we see that the continuum limit only changes these energy--momentum
relations to the standard relativistic form,
$(E_j,\,P_j)=m(\cosh\t_j,\,\sinh\t_j)$, without affecting the $S-$matrix
as a function of $\t$.

\chapter{SOS and RSOS reductions and kink interpretation}

Under the vertex--face correspondence previously described, the light-cone
6V model is mapped into the SOS model. At the level of the Hilbert space, this
corresponds to the restriction to the highest weight states of $SU(2)_q$:
each $(2J+1)-$dimensional multiplet of spin $J$ (for generic, non--rational
values of $\g/\pi$) is regarded as a single state of the SOS model. In
particular, in such a state the local height variables $\ell_n$ have
well defined boundary values $\ell_0=1$ and $\ell_N=2J+1$. The ground state
of the 6V model, which is a $SU(2)_q$ singlet, is also the ground state of the
SOS model: it has $\ell_0=\ell_N=1$ and is ``dominated", in the thermodinamic
limit $N\to\infty$ by the see-saw configuration depicted in fig. 1. Our
boundary conditions allow for only one such ``ground state dominating"
configuration, with  $\ell_n=(3-(-)^n)/2$, while periodic b.c. {\it on the SOS
variables $\ell_n$}  would allow any possibility: $\ell_n=(2\ell+1\pm(-)^n)/2$,
with $\ell$ any positive integer.

Consider now a BA state with one hole. Then $N$ is necessarily odd while
$J=1/2$ and $\ell_N=2J+1=2$. From the SOS point of view this state is
``dominated" by height configurations of the type depicted in fig.4. Upon
``renormalization", we can replace the ground state and one hole configurations
with the smoothed ones of fig. 5. The hole corresponds, in configuration space,
to a kink of the SOS model which interpolates between two neighboring vacuum
states: the vacuum on the left of the kink has $\ell_n=(3-(-)^n)/2$,
corresponding to a constant ``renormalized" height ${\hat\ell}_n=1$, while the
vacuum on the right has $\ell_n=(5+(-)^n)/2$, corresponding to
${\hat\ell}_n=2$. Of course, many kink configurations like that depicted in
fig. 4 are to be combined into a standing wave (a plane wave in the infinite
volume limit which turns the one hole BA state into  an eigenstate of
momentum). Let us observe that, by expressions like ``dominating
configuration", we do not mean that, e.g. the ground state, becomes an
eigenstate of $\ell_n$ as $N\to\infty$. We expect local height fluctuations
to be present even in the thermodynamic limit or, in other words, that the
ground state remains a superpositions of different height configurations.
The identification of a dominating configuration is made possible by the
integrability of the model, which guarantees the existence of the higher--level
BA. In turns, the higher--level BA allows us to consistently interpret the
ground state or the one hole state as in figs. 1, 4 and 5, since the presence
of fluctuations only renormalizes in a trivial way the scattering of physical
excitations relative to the bare, or microscopic, $R-$matrix \rma, as evident
from eq.\sma. Therefore, we can reinterpret at the renormalized level the
vertex-face correspondence: the holes, that is the solitons of the SG model,
are $SU(2)_q$ doublets acting as SOS kinks that increase or decrease by 1 the
renormalized heights ${\hat\ell}_n$. The higher--level BA makes sure that the
total number of internal states of $\nu$ kinks interpolating between
${\hat\ell}=1$ and ${\hat\ell}=2J$ is just the number of highest weight states
of spin $J$ in the tensor product of $\nu$ doublets
$$
       d_\nu(J)={\nu\choose{\hat M}} - {\nu\choose{{\hat M}-1}}
             \quad ;\quad {\hat M}=[\nu/2]-J               \eqn\nkink
$$
This parallels exactly the original BA, which provides the
$$
       d_N(J)={N\choose M} - {N\choose{M-1}}
             \quad ;\quad {\hat M}=[N/2]-J                 \eqn\nhws
$$
highest wight states of $N$ doublets with total spin $J$.

Comparing eqs. \rma, \peso\ and \sma, we can directly write down the $S-$matrix
of the SOS kinks:
$$
     S(\t)^{is,sj}_{ir,rj}=S_0(\t)\left\{
            \delta_{rs}+\delta_{ij}\,b({\hat\t},{\hat\g})
    \left[{\sin{\hat\g} r\,\sin{\hat\g} s}
         \o{\sin{\hat\g} i\, \sin{\hat\g} j}\right]^{1/2}
                         \right\}                            \eqn\skink
$$
It defines the two--body scattering as follows. The first ingoing kink
interpolates between the local vacuum with ${\hat\ell}=i$ and that with
${\hat\ell}=r\;\;(r=i\pm 1)$, while the second interpolates between
${\hat\ell}=r$ and ${\hat\ell}=j\;\;(j=r\pm 1)$. The outgoing kinks are
interpolating between ${\hat\ell}=i$ and ${\hat\ell}=s\;\;(s=i\pm 1, i.e,
s=r,r\pm 2)$ and between ${\hat\ell}=s$ and ${\hat\ell}=j\;\;(j=s\pm 1)$.

When $\g=\pi/p$ for $p=3,4,\ldots$, the SOS models can be restricted to the
RSOS($p$) models, by imposing $\ell_n<p$. In the vertex language of the 6V
model
this corresponds to the Hilbert space reduction to the subspace formed by the
so--called type II representations of $SU(2)_q$ with $q^p=-1$ [\qru
,\qrud ]. In the
next section we shall describe in detail how the restriction takes place in our
BA framework. Here we simply observe that the local height restriction, when
combined with the finite renormalization $\g\to{\hat\g}$ of eq.\ren,
provides a strong support for the kink interpretation presented above.
Indeed, if $\g=\pi/p$, then ${\hat\g}=\pi/(p-1)$ and the renormalized heights
${\hat\ell}_n$ can take the values $1,2,\ldots,p-2$. This appears now obvious,
since each constant configuration of ${\hat\ell}_n$ corresponds to an $\ell_n-$
configuration oscillating two neighboring values. Moreover under the standard
identification of the critical RSOS($p$) models with the minimal CFT series
$M_p$, we see that each light--cone RSOS($p$) model has kink excitations
whose $S-$matrix \skink\ is proportional to the (complex) microscopic Boltzmann
wights of the RSOS($p-1$) model, under the replacement of $2\Th=\Th-(-\Th)$=
rapidity difference of light-cone right and left movers, with $(p-1)$ times
$\t=\t_1-\t_2$ =rapidity difference of physical particles. This is exactly the
pattern found by bootstrap techniques [\many] for the minimal model $M_p$
perturbed by the primary operator $\phi_{1.3}$ (with negative coupling).

\chapter{BA roots when $q$ is a root of unity and Quantum Group reduction.}

As previously explained to each $M$ roots solution of the BAE there corresponds
a highest weight state of the quantum group $SU(2)_q$ with spin $J=N/2-M$.
It is well known that when $q$ is a root of unity, say $q^p=\pm 1$, then
$(J_+)^p=(J_-)^p=0$ and the representations of $SU(2)_q$ divide into two very
different types. Type I representations are reducible and generally
indecomposable. They can be described as pairwise mixings of standard irreps
(that is the irreducible representations for $q$ not a root of unity) with spin
$J$ and $J^\prime$ such that
$$
    |J-J^\prime|<p  \;,\qquad   J+J^\prime=p-1 \;\;({\rm mod\,} p) \eqn\mixrule
$$
Notice that the sum of $q-$dimensions for this pair of reps vanishes.

Type II representations are all the others. They are still fully irreducible
and structured just like the usual $SU(2)$ irreps. Since $(J_\pm)^p=0$,
type II representations have necessarily dimension smaller than $p$, that is
$$
                  J < {{p-1}\o 2}                              \eqn\bound
$$
\indent We shall now show how the BAE reflect these peculiar properties of the
$SU(2)_q$ representations for $q$ a root of unity. First of all let us stress
that when $q$ is not a root of unity (\ie $\g/\pi$ is irrational), then the BAE
cannot possess $v-$roots at (real) infinity. Indeed, since $\B(\infty)$ is
proportional to $J_-$ (eq.\bj), a root at infinity means that the
corresponding BA state is obtained by the action of the lowering operator $J_-$
on some other state with higher spin projection $J_z$. But for $q$ not a root
of unity, the BA states have $J=J_z$ and therefore cannot be obtained by
applying $J_-$ on any other state. On the other hand, if we assume that one
$z-$root, say $z_1$, of the BAE \baet\ lay at the origin,
\ie $Re\,v_1=+\infty$,
then eqs. \baet\ for  $j=1$ imply
$$
                        q^{4(J+1)} =1                             \eqn\qru
$$
That is, $q$ must be a root of unity. It is now crucial to observe that the
remaining equations for the non--zero roots (those labelled by
$j=2,3,\ldots,M$) are precisely the BAE for $M-1$ unknowns. This invariance
property holds only for the quantum group covariant, fixed boundary conditions
BAE. It does not hold for the p.b.c. BAE where $v-$roots at infinity twist the
remaining equations for finite roots. This twisting reflect the fact that the
corresponding periodic row--to--row tranfer matrix is not quantum group
invariant but gets twisted under $SU(2)_q$ transformations [\rev ].
Thus, when $q$
is a root of unity, BA states \abald\ with one $v-$root at infinity take the
form
$$
          \Psi(v_1=\infty,v_2,\ldots,\sub vM)=
                 (1-q^2)J_-\Psi(v_2,v_3,\ldots,\sub vM)  \eqn\bavecr
$$
Let us recall that the BA state on the l.h.s. is annihilated by $J_+$ for any
$q$ (including $q$ a root of unity) [\ddvI]. Hence eq.\bavecr\ represents the
mixing of two reps with spin $J$ and $J^\prime=J+1$ into a type I
representation. Indeed, applying the mixing rule \mixrule\ into the necessary
condition \qru\ for one root at infinity, yields an identity as required:
$$
       1=q^{4(J+1)}=q^{2(J+J^\prime+1)}=\left(q^p \right)^{2n}=1      \eqn\ide
$$
where $n$ is a suitable ($J-$ dependent) positive integer.

We can generalize this analysis to any number of vanishing BA $z-$roots. Let us
identify the roots going to the origin with $z_1,z_2,\ldots,z_r$,
$1\le r\le M$. The BAE for the remaining non--zero roots
$z_{r+1},\ldots,\sub zM$
take the standard form \baet\ valid for a BA state formed by $M-r$
$\B-$operators, which has therefore spin $J+r$. The BAE for the vanishing roots
take the form
$$
    q^{4J+2(r+1)}=F_j \equiv
     \prod_{\scriptstyle k=1\atop\scriptstyle k\ne j}^r
          {{z_j q^2-z_k}\o{z_j-z_k q^2}} \;,\quad 1\le j\le r      \eqn\vbae
$$
where the definition of $F_j$ is understood in the limit of vanishing
$z_1,\ldots,z_r$. We can obtain complete agreement with the quantum group
mixing rules \mixrule\ by setting $r<p$ and making the natural choice
$$
            F_j=1\;,\quad   1\le j \le r                         \eqn\natural
$$
Indeed the presence of $r$ vanishing  $z-$roots imply the mixing of two
representations with spin $J$ and $J^\prime=J+r$, so that
$$
       q^{4J+2(r+1)}= q^{2(J+J^\prime+1)}= \left(q\right)^{2n}=1  \eqn\agree
$$
with $n$ an integer depending on $J$ and $J^\prime$. Eq. \natural\ ha a very
simple solution for the limiting behaviour of the vanishing roots. We find that
if $$
              z_j=\omega^{j-1}\,z_1 \;,\quad   1\le j \le r    \eqn\limit
$$
with $\omega^r=1$, then identically
$$
      F_j=\prod_{\scriptstyle k=1\atop\scriptstyle k\ne j}^r
         {{\sin[-\g+\pi(k-j)/r]}\o{\sin[\g+\pi(k-j)/r]}}=1
             \;,\quad 1\le j\le r                             \eqn\ident
$$
for any value of $\g$. That eq.\ident\ should hold for generic values of
$q=e^{i\g}$ is necessary since we are studying the approach of the roots to the
origin when $\g/\pi$ tends to a rational number. We would like to remark that
the limiting behaviour of the ratios of the vanishing BA $z-$roots depend only
on their number $r$ and neither on the specific rational value of $\g/\pi$ nor
on the other non--vanishing roots.

In the $v-$plane, the roots $v_1,v_2,\ldots,v_r$ go to infinity as a
$r-$string with spacing $i\pi/r$. For example, when $r=2$ we have a pair of
complex roots with imaginary parts tending to $\pm i\pi/4$ as their real parts
simultaneously diverge. For $r=3$ there will be a limiting 3-string with a real
member and a complex pair at $\pm i\pi/3$. For these two examples, we have
performed an explicit numerical test when $\g\to(\pi/3)^-$ and
$\g\to(\pi/4)^-$, respectively, finding perfect agreement with the picture
proposed here. We shall present more detail on the special cases in the next
section. Let us anticipate here that the numerical results suggest the
following conjecture on the behaviour of the singular roots:
$|z_j|\approx O(\e^{1/r})$ and  ${\rm Im}[v_j-v_j(\e=0)]\approx O(\e^{1/r})$
where $\e=\g-\pi/(r+1)$.

When $q^p=\pm 1$ and the $SU(2)_q$ reps divide into type I and type II, it is
possible to perform the consistent RSOS($p$) reduction of the Hilbert space.
This consists in keeping, among all states annihilated by $J_+$ which form the
SOS  subspace, only the type II states. It is known that this corresponds to
SOS configurations with local heights $\ell_n$ restricted to the set
${1,2,\ldots,p-1}$ [\qru \qrud]. In our BA framework, therefore, the RSOS
reduction
is obtained by retaining only those states with spin $J<(p-1)/2$, that is with
$M>(N-p+1)/2$, such that all BA roots $z_1,z_2,\ldots,z_r$ are non--zero. This
constitues the general and simple prescription to select the RSOS($p$) subspace
of eigenstates of the evolution operator when $q^p=\pm 1$.

We have considered up to here the RSOS reduction at the microscopic level, that
is for the BA states \abald\ described by the bare BAE \baet. It should be
clear, however, that the analysis of the singular BAE solutions when $q$ is a
root of unity holds equally well for the higher--level BAE \hbae, due to their
structural identity with the bare BAE. The  crucial problem is whether a
quantum group reduction carried out the higher level would be equivalent to
that described above in the ``bare" framework. This equivalence can actually be
established as follows. Suppose that $\g/\pi$ is irrational, but as close as we
like to a specific rational value. Then all solutions of the bare BAE are
regular and can be correctly analyzed, in the limit $N\to\infty$, using the
density description for the real roots. As described in paper I, this yields
the higher--level BAE \hbae, those roots are  in a precise correspondance with
the {\it complex} roots of the bare BAE. When $\g$ tends to $\pi/p$, then
${\hat\g}$ tends to $\pi/(p-1)$ and  singular solutions of the bare BAE with
two
or more singular roots  are in one--to--one correspondance with singular
solutions of the higher--level BAE. Thus in this case quantum group reduction
and renormalization of the BAE commute.
The only potentially troublesome cases are those of singular solutions with
only one real singular root. Indeed the density description of the $v-$roots
cannot account, by construction, for {\it real} roots at infinity, and no sign
of the type I nature of the corresponding BA state would show up at the higher
level. Notice that spin $J$ BA states with a single singular  root are mixing
with spin $J+1$ states, so that, by the quantum group mixing rule \mixrule,
necessarily $J=p/2-1$.  On the other hand, a direct application of the
constraint \bound\ at the higher level, that is with the replacement
$p\to {\hat p}=p-1$, would overdo the job, by incorrectly ruling out all BA
states with $J=p/2-1=({\hat p}-1)/2$. To be
definite, consider $N$ even. Then the type II BA states with $J=p/2-1$, which
ar
e
superpositions of SOS configurations with $\ell_n\le p-1=2J+1=\ell_N$, have
an effective, higher--level spin ${\hat J}\equiv({\hat\ell}_N-1)/2= J-1/2$.
This
higher--level spin does satisfy \bound. The type I states are those in which
$\ell_n$ somewhere exceeds $p-1$. In particular, the simplest change on a type
II configuration, turning it into type I, is to flip the oscillating $\ell_n$
close to the right wall (as shown in fig. 6). $\ell_N$ and hence the spin $J$
are left unchanged, but clearly ${\hat\ell}_N$ increases by one, causing
${\hat J}$ to violate the bound \bound. In other  words, RSOS($p)-$acceptable
BA
states with $J=p/2-1$ have one kink less than the corresponding SOS states.
 From the detailed analysis of the $p=4$ case, to be discussed below, it
appears
that the removal of one kink corresponds to giving infinite rapidity to the
hole representing that kink in the higher--level BAE.

Thank to the possibility of performing the restriction directly at the
renormalized level, the kink $S-$matrix of the RSOS($p$) model follows by
direct  restriction on the SOS $S-$matrix given by eq.\skink. Namely, one must
set $\g=\pi/p$, that is  ${\hat\g}=\pi/(p-1)$, and consider all indices as
running from 1 to $p-2$.

\chapter{The models RSOS(3) and RSOS(4)}

In order to clarify matters about the BA RSOS reduction discussed in the
previous section, we present here the more details about the two simplest
examples, when $\g=\pi/3$ and $\g=\pi/4$. For these two cases the RSOS
reductions  correspond, respectively, to a trivial one--state model and to the
Ising model (at zero external field and non--critical temperature). Let us
recall that the light--cone approach yields in the continuum limit massive
field theories with the same internal symmetry of the corresponding critical
regimes. Therefore the RSOS(4) reduction of the light--cone 6V model coincides
with the $\ZZ_2-$ preserving perturbation of the $c=1/2$ minimal model (which
is
obtained upon quantum group restriction of the critical 6V model).

\section{The case $p=3$}
{}From the RSOS viewpoint, the case $\g=\pi/3$ is particulary simple. Since
$p=3$
the local hight variables $\ell_n$ can assume only the values 1 and 2. Then the
SOS adjacency rule $|\ell-\ell^\prime|=1$ implies that the global
configuration is completely determined once the height of any given site
is chosen. In our light--cone formulation the first height on the left is
frozen to the value 1 (see eq.\locl\ and fig. 2), so that the restricted
Hilbert space contains only one state: an $SU(2)_q$ singlet when $N$ is even
and the spin up component of a $J=1/2$ doublet when $N$ is odd.

For $\g=\pi/3$ the BAE still have many solutions which reproduce the full SOS
Hilbert space. Indeed, from the SOS point of view, nothing particular happens
when $\g\to\pi/3$. However, at this precise value of the anisptropy, only one
BAE solution corresponds to the unique type II representation: for $N$ even
(odd) it contains $N/2\,(N/2-1/2)$ non--zero roots in the $z-$plane. All other
BAE solutions with $M=[N/2]$ contain at least one vanishing
$z-$root and correspond to type I SOS states. Moreover, there unique RSOS state
is the ground state of the SOS model and is formed by real positive roots
labelled by consecutive quantum integers (no holes). We reach therefore the
following rather non--trivial conclusion: the complicated system of algebraic
equations \baet\ admit, for $q^3=\pm1$ and $w$ real, one and only one solution
with $M=[N/2]$ non--zero roots within the unit circle. In addition, these roots
are real and positive. For few specific choices of $N$ we also verified
this picture numerically.

\section{The case $p=4$}

The local height variables can assume now the three values $\ell=1,2,3$.
However, each configuration can be decomposed into two sub--configurations
laying on the two sublattices formed by even and odd faces, respectively.
On one of the two, the SOS adjacency rule $|\ell-\ell^\prime|=1$ freezes the
local heights to take the constant value $\ell=2$. Then on the other
sub--lattice we are left with two possibilities, $\ell=1$ and 3. Moreover,
the interaction round--a--site of the original model reduces in this
way to a nearest--neighbor interaction in the vertical and horizontal
directions. The framework is that of the Ising model.

To obtain the standard Ising formulation, we can set
$$
               \s_n(t)=\ell_{2n}(t)-2              \eqn\defis
$$
where the numbering of the lattice faces can be read from fig. 2.
The fixed b.c. on the $\ell_n$ now correspond to
$$\eqalign{
  \s_0(t)&=-1 \;,\quad \sub{\s}R(t)=(-1)^{J-1} \qquad (N=2R)\cr
  \s_0(t)&=-1 \;,\quad \sub{\s}R(t)=\pm 1 \qquad\quad (N=2R+1)\cr}   \eqn\bcs
$$
where $J=0,1$ for $N$ even, due to the bound \bound. For $N$ odd we must
consider only the possibility $J=1/2$, which implies $\ell_N=2$, since
the line of half--plaquettes on the extreme right belong to the frozen
sublattice. Then $\ell_{N-1}$ is left free to fluctuate between 1 and 3,
leading to free b.c. on $\s$. The matrix elements of the unit time evolution
operator ${\tilde U}$ can be calculated from the
explicit form of the SOS weights \dobw -\peso. They can be written in the
``lagrangian" form
$$
              \sub{\vphantom{\big(\big)}}{t+1}
             \bra{\s_0^\prime,\s_1^\prime,\ldots \s_R^\prime} {\tilde U}
     \ket{\s_0,\s_1,\ldots \sub{\s}R}_t = e^{iL(t)}                    \eqn\act
$$
where
$$
    L(t)=\b_v\sum_{n=1}^{R-1} \s_n(t)\s_n(t+1)
        -\b_h\sum_{n=0}^{R-1} \s_n(t)\s_{n+1}(t)+ {\rm const}        \eqn\lagr
$$
and $$
      \b_v=\frac14(\pi+2i\ln\tanh 2\Th)\;,\quad
                \b_h=\arctan\tanh 2\Th                            \eqn\bet
$$
Alternatively, standard simple manipulations allow to rewrite ${\tilde U}$
explicitly in terms of Pauli matrices
$$\eqalign{
          {\tilde U}=&e^{-i\b_h H_2} e^{-i\b_h H_1}  \cr
          H_1=&\sum_{n=1}^{R-1}(\s_n^x+1) \;,\quad
          H_2=\sum_{n=0}^{R-1}\s_n^z\s_{n+1}^z   \cr}             \eqn\pauli
$$
In either cases, one sees that the {\it complex} Boltzmann weights {\it
formally}  belong to the critical line
$$
              \sin 2\b_v \sin 2\b_h =1                            \eqn\crit
$$
Of course, this follows from the original definition of the light--cone 6V
model in terms of complex trigonometric Boltzmann weights, which, under the
replacement $\Th\to i\Th$, would correspond to the critical standard 6V model.
Nevertheless, just as for the vertex model, also in this ``light--cone" Ising
model, a massive field theory can be constructed in the
${\rm Re\,}\Th\to +\infty$ limit.

In the BA diagonalization of ${\tilde U}$, we must restrict ourselves to the
BAE solutions with $M=R$ or $M=R-1$ for even $N$ and $M=R$ for odd $N$.
For sufficently large even $N$, the ground state has $R$ real roots with
consecutive quantum integers. Actually, as already stated above,
this is a general fact valid for all RSOS($p$) models. Namely, the infinite
volume ground state of th SG model, of the SOS model and of all its
restrictions RSOS($p$) is the same f.b.c. BA state. It is the unique $SU(2)_q$
singlet with all real positive roots and no holes. It is described in the
thermodynamic limit ($N\to\infty$, $a$ fixed) by the density of roots given in
eq.(5.7) of paper I.

Excited states with $J=0$ have an even number of holes and a certain number of
complex roots such that no $z-$root lays at the origin. For instance a
two--particle state contains two holes (holes are naturally identified with the
particles) and a two--string with imaginary parts $\pm[\pi/8+$ corrections
exponentially small in $N]$ laying between the two holes. This state (that is
this precise choice of quantum integers) is just the two--particle state of the
SG or SOS model, with $\g$ fixed to the precise value $\pi/4$. We have
explicitly checked, by numerically solving the BAE for various values of $N$,
that indeed all $z-$roots stay away from the origin as $\g$ crosses $\pi/4$
while the quantum integers are kept fixed to the two--hole configuration.
Now consider a state with four holes. Apart from the multiplicity of the
rapidity phase space, there are two distinct type of such states: in $v-$space
one contains two two--strings, while the other contains one sigle wide pair,
that is a complex pair with imaginary part larger than $\g$. This situation
holds for generic values of $\g/\pi$ and simply reflects the fact that the
holes
are  $SU(2)_q$ doublets. The crucial point is that, as $\g$ reaches
$\pi/4$ (from below), the wide pair moves towards infinity, while its   real
part gets closer and closer to the (diverging) value of the  largest real root
and the imaginary part approaches $\pi/3$. This picture is confirmed by a
careful numerical study of the BAE and is in perfect agreement with the general
picture presented in sec. 5. In addition, there exists numerical evidence that
the real parts diverge like $\log(\pi/4-\g)^{-1/6}$ while the imaginary part of
the wide pair goes to $\pi/3$ like $(\pi/4-\g)^{1/3}$.
When $\g>\pi/4$, the $v-$root corresponding to the
largest quantum integer has the largest real part, but is  no longer real.
having an imaginary part equal to $\pi/2$. As $\g\to(\pi/4)^+$, this real part
again diverges together with real part of the wide pair. Hence the four--hole
state with a wide pair is type I when $\g=\pi/4$. On the other hand, the
four--hole state with two two--string is type II, since all its $v-$roots stay
finite.

 From the study of the two-- and four--hole states, we are led to the following
general conjecture: in the BA framework, the RSOS(4) $J=0$ states are all and
only the states with  $\nu=2k$ holes and $k$ two--strings. Then, in the
higher--level BAE \hbae, we recognize this state as that corresponding to the
unique solution with $k$ real  $\chi-$roots. In other words, this BA state is
completely determined once the location of the holes (that is the rapidities of
the physical particles) is given.

Consider now the states with $J=1$. In this sector, the lowest energy
type II state contains exactly one hole. This is an unusual situation for BA
systems on lattices with even $N$, where holes are always treated in pairs.
As long as $\g\not=\pi/4$, the same is true in our f.b.c. BA: the lowest energy
state with $J=1$ contains two holes, in agreement with the interpretation of
holes as SG solitons with quantum spin 1/2. For $\g<\pi/4$ all roots are real.
For $\g>\pi/4$ the root $v_{N/2-1}$ corresponding to the largest quantum
integer $I_{N/2-1}=N/2+1$ aquires an imaginary part $\pi/2$ (see the appendix
for details). But when  $\g=\pi/4$ then Re$\,v_{N/2-1}=+\infty$, and the
$J=1$ two--hole state is mixed with some $J=2$ state into a type I
representation. It does not belong to the RSOS(4) Hilbert space.
This picture can be easily verified numerically.

To prevent the largest root $v_{N/2-1}$ from diverging, it is sufficent to
consider a $J=1$ state with only one hole and $I_{N/2-1}=N/2$. From the point
of view of the SG or SOS models this one--hole state has a cutoff--dependent
energy which diverges as $a^{-1}$ in the continuum limit. If $x_1$ is the
position of the single hole, the energy relative to the ground state reads
$$
          E-E_0= a^{-1}{\rm e}(x_1) +\pi a^{-1}             \eqn\diven
$$
where the renormalized energy function ${\rm e}(x)$ is given in eq.\enhole.
Notice that this result coincides with the limit $x_2\to\infty$ of a two--hole
state (cft. eq.\enhole). Strictly speaking, however, the density method
leading to \enhole\ has no justification when ``one hole is at infinity".
We trust eq.\diven\ nonetheless because it passes all our numerical checks.
In the continuum limit $a\to0,\;\Th\to\infty,\,4a^{-1}e^{-\pi\Th/\g}=m$
(fixed), this one--hole state is removed from the physical spectrum, as
required from the SG point of view: holes are spin 1/2 solitons while here
$J=1$.

The preceding discussion easily extends to generic states in the $J=1$ sector
containing an odd number of holes and a given set of complex pairs. The
infinite--volume energy of these multiparticle states is given by
$$
    E=E_0+a^{-1}\sum_{h=1}^\nu {\rm e}(x_h)+\pi a^{-1}   \eqn\mdiven
$$
with the same divergent constant $\pi a^{-1}$ appearing irrespective of the
physical content of the state. Our conjecture for the $J=0$ states naturally
extends to these $J=1$ states: if there are $\nu=2k+1$ holes, then the
$v-$roots are all finite, implying that the state is type II, provided there
are also $k$ two-strings. Then the higher--level BAE imply that these states
are completely defined by the hole rapidities. To retain these type II $J=1$
BA states in the RSOS(4) model, an extra $J-$dependent subtraction is necessary
to get rid of the divergent constant. Namely, for $N$ even and $J=0$ or 1, we
set $$
           H_{RSOS(4)}=H_{SOS}(\g=\pi/4) - \pi a^{-1}J            \eqn\subt
$$
where $H_{SOS}$ is given by eqs. \uop, \hami, \gugm -\peso .

Alltogether, we see that the BA picture for the excitations of the RSOS(4)
model is fully consistent with the kink interpretation for the holes.
Indeed $J=0$ corresponds to even Dirichlet b.c. for the Ising field, while
$J=1$ corresponds to odd Dirichlet b.c. (cft. eqs. \bcs).
In the Ising model the kink description is valid below the critical
temperature, with the disorder field as natural interpolator for the kinks.
In this case the $S-$matrix must be $-1$ (it is $+1$ when the asymptotic
particles are the free massive Majorana fermions) and this is exactly what
follows from eq.\skink\ upon setting ${\hat\g}=\pi/3$ and restricting
all indices to run from 1 to $p-2=2$.

The analisys of the BA spectrum for odd $N$ does not contain real new
features. Now $J$ is fixed to 1/2 and the b.c. on the Ising field are of mixed
fixed--free type, as  shown in the second of eqs. \bcs. The lowest energy state
corresponds to the  BAE solution formed by real roots and one single hole as
close as possible to the the $v-$origin. In other words, the quantum integers
are given by $I_j=j+1$, for $j=1,2,\ldots,(N-1)/2$. By letting this hole to
move away along the positive real axis, we reconstruct the energy spectrum
of a one--particle state. Of course, to compare this state to the global ground
state, which contains no holes, a judicious choice of the odd value of $N$ is
required. If in the ground state $N=2R$, then the new state is indeed an
excited state if we choose $N=2R-1$, rather than $N=2R+1$, because of the
antiferromagnetic nature of the interaction.

\appendix

In this appendix, we present for completeness a (rather non--standard)
treatment of the BAE \baet. As explained in paper I, it is convenient to first
rewrite them in the p.b.c. form
$$\eqalign{
     \left[{{\sinh(\l_j-\Th+i\g/2)\,\sinh(\l_j+\Th+i\g/2)}
          \o{\sinh(\l_j-\Th-i\g/2)\,\sinh(\l_j+\Th-i\g/2)}}\right]^N
       &\, {{\sinh(2\l_j+i\g)}\o{\sinh(2\l_j-i\g)}}\cr =-\prod_{k=-M+1}
          {{\sinh(\l_j-\l_k+i\g)}\o{\sinh(\l_j-\l_k-i\g)}}\cr}    \eqn\pbcform
$$
where the numbers $\sub{\l}{-M+1},\sub{\l}{-M+2},\ldots,\sub\l M$ are related
to
the $v-$roots by
$$
         \l_j=v_j=-\l_{1-j} \;,\quad  j=1,2,\ldots,M      \eqn\vtol
$$
and ${\rm Re\,}v_j>0$, thanks to the symmetries of the BAE \baet. Let us
concentrate our attention on those BAE solutions which are mostly real, that is
those which contain an arbitrary but fixed number of complex pairs interspersed
in a sea of order $N$ of real roots. The reason for this restriction will
become clear in the sequel. The {\it counting function} associated to a given
solution $v_1,\ldots,\sub vM$ is defined to be
$$\eqalign{
      Z_N(x\,;v_1,\ldots,\sub vM)=&\phi_{\g/2}(x+\Th)+\phi_{\g/2}(x-\Th)
                                +N^{-1}\phi_\g(2x)    \cr
          -& N^{-1}\sum_{j=-M+1} \phi_\g(x-\l_j)    \cr}        \eqn\count
$$
where $$
            \phi_\a(x) \equiv i\log{{\sinh(i\a+x)}\o{\sinh(i\a-x)}} \eqn\deffi
$$
The logarithmic branch in eq.\deffi\ is chosen such that $\phi_\a(x)$, and
as direct consequence $Z_N(x\,;v_1,\ldots,\sub vM)$, are odd.

The BAE \pbcform\ can now be written in compact form
$$
       Z_N(\l_j\,;v_1,\ldots,\sub vM)= 2\pi N^{-1}\,I_j
                           \;,\quad  j=-M+1,-M=2,\ldots,M      \eqn\compact
$$
where the {\it quantum integers} $I_j$ entirely fix the specific BAE solution,
and therefore the BA eigenstate, and by construction satisfy $I_{1-j}=-I_j$ for
$j=1,2,\ldots,M$. Notice also that by definition
$Z_N(0\,;v_1,\ldots,\sub vM)$=0, but $\l=0$ is not a root, due to \vtol.
Rather, $\l=0$ is always a {\it hole}, from the p.b.c. point of view.

To begin, consider the case when $N$ is even, $M=N/2$ (\ie\ $J=0$) and all
roots are real. This state (the groud state for even $N$) is unambiguously
identified by the quantum integers
$$
                I_j=j \quad\;\qquad    j=1,2,\ldots,N/2            \eqn\gs
$$
and the corresponding counting function is indeed monotonically
increasing on the real axis, justifying its name. To our knowledge, the
existence itself of this BAE solution has not been proven in a rigorous
analytical way.
But it is very easy to obtain it numerically for values of $N$ in the thousands
and precisions of order $10^{-15}$ on any common workstation.

Next consider removing $J$ roots from the ground state. $J$ is the quantum spin
of the corresponding new BA state. For $\g$ sufficently small, one now finds
that$$
        N/2+J < Z_N(+\infty\,;v_1,\ldots,\sub vM)<N/2+J+1     \eqn\asym
$$
so that, together with the actual roots satisfying eq.\compact, there must
exist positive numbers $x_1,x_2,\ldots,x_\nu$, with $\nu\ge 2J$, satisfying
$$
       Z_N(x_h\,;v_1,\ldots,\sub vM)= 2\pi N^{-1}\,{\bar I}_h
                           \;,\quad  h=1,2,\ldots,\nu        \eqn\hole
$$
where the ${\bar I}_h$ are positive integers. If $Z_N(x)$ is monotonically
increasing, then necessarily $\nu=2J$ and the integers
$\{I_1,\ldots,I_{N/2-J},\,{\bar I}_1,\ldots,{\bar I}_{2J}\}$ are all distinct.
The numbers $x_1,x_2,\ldots,x_\nu$ are naturally called {\it holes}.
For $J$ held fixed as $N$ becomes larger and larger, it is natural to expect
that the counting function is indeed monotonically increasing, and numerical
calculations confirm this expectation. For larger values of $\g$ the situation
becomes more involved. Numerical studies show that, first of all, $Z_n(x)$
develops a local maximum beyond the largest root, while still satisfying the
bounds \asym, as $\g$ exceeds a certain ($J-$dependent) value. For even larger
values of $\g$, the asymptotic value of $Z_N(x)$ becomes smaller than
$Z^\ast\equiv 2\pi(1/2+J/N)$,
but its maximum stays larger than $Z^\ast$, provided there is indeed a
root  corresponding to $N/2+J$ (\ie\ $I_M=N/2+J$). Up to now,
$\sub vM$ is obviously located where $Z_N(x)$ reaches $N/2+J$ from below, and
one could say that there exists an extra hole $\sub{x}{2J+1}$ further beyond,
where $Z_N(x)$ reaches  $Z^\ast$ from above. When $\g$ reaches a certain
critical  value, the local maximum lowers till $Z^\ast$. At this point the root
and the extra hole exchange their places, and for sligtly larger values of $\g$
the hole is located where $Z_N(x)$ reaches $Z^\ast$ from below, while the root
lays  further away, where $Z_N(x)$ reaches  $Z^\ast$ from above. As the
numerical  calculations show,  however, this extra hole with the same quantum
integer $I_M=N/2+J$ of the largest root  hole is spurious, since no  energy
increase is
really associated to its presence. When none of $N/2-J$ root has $N/2+J$ as
quantum integer, then $Z_N(x)$ does never reach $Z^\ast$ for sufficently large
$\g$, and, strictly speaking there are only $2J-1$ holes. This time, however,
we find that the energy increases with respect to the ground
state in the same way
as if there was ``a hole at infinity", that is a hole beyond the largest root.
Thus $\nu$ can always be regarded to be even, when $N$ is even.

As an important example consider the definite choice $J=1$. Then for
$\g<\pi/6$ we find $Z_N(+\infty)> Z^\ast$, and there are
two holes, with
$1\le{\bar I}_1<{\bar I}_2\le N/2+1$. Assume ${\bar I}_2\le N/2+1$ and consider
the interval  $\pi/6<\g<\pi/4$. The counting function has a maximum
$Z_{max}$
(larger than $Z^\ast$) situated to the right of the largest root
$\sub v{N/1-1}$,
as long as $\g$ is smaller than the critical value $\g^\ast$  at which the
maximum
lowers to $Z^\ast$. For $\g>\g^\ast$, the maximum is still larger than $Z^\ast$
 but
is located to the left of $\sub v{N/1-1}$. In any case, for $\pi/6<\g<\pi/4$
the asymptotic value
$Z_N(+\infty)$ is smaller than $Z^\ast$. When $\g\to(\pi/4)^-$, then the
largest root as well as the maximum $Z_{max}$ are pushed to infinity, and
$Z_N(x)$ is once again monotonic with $Z_N(+\infty)=Z^\ast$. As $\g$ exceeds
$\pi/4$, the last root $\sub v{N/1-1}$ passes, through the point at infinity,
from the real line to the line with ${\rm Im}\,v=\pi/2$. This pictures
generalizes to arbitrary $J$ with the two special values $\g=\pi/6$ and
$\g=\pi/4$ replaced, respectively, by  $\g=\pi/(4J+2)$ and $\g=\pi/(2J+2)$.
In fig. 7 the salient portion of the numerically calculated counting function
is plotted for $J=1$, $N=64$ and a specific choice of
${\bar I}_1,\;{\bar I}_2$. In this case we approximatively find $\g^\ast\simeq
0.21\pi$.

Finally, consider a BAE solution containing, in addition to a number of order
$N$ of real roots, also a certain configuration of complex roots. In the
$v-$space, these complex roots appear either in complex conjugate pairs or with
fixed imaginary part equal to $i\pi/2$, so that the counting function is real
analytic: ${\overline{Z_N(x)}}=Z_N({\bar x})$. Moreover, it is fairly easy to
show, by looking at the value of the counting function a real infinity, that
the presence of complex roots implies the existence of holes in the sea of real
roots. For our next porposes, we shall now consider $\g/\pi$ irrational, so
that all $v-$roots are finite. Denoting with $u_q,\;q=1,2,\ldots,M_c$ the
values of the complex roots and with $M_r=M-M_c$ the number of real roots, we
now write the derivative of the counting function as
$$
            2\pi\rho_N(x)\equiv Z_N^\prime(x)=F_\Th^\prime(x)+
        N^{-1}\left[F_0^\prime(x)+F_c^\prime(x)-F_h^\prime(x)\right]+
                    (K\ast \rho_\delta)(x)                         \eqn\deri
$$
where $$\eqalign{
  F_\Th(x)=&\phi_{\g/2}(x+\Th)+\phi_{\g/2}(x-\Th)        \cr
  F_0(x)=  &\phi_\g(2x)                                 \cr
  F_c(x)=  &\sum_{q=1}^{M_c}\left[\phi_\g(x-u_q)+\phi_\g(x+u_q)\right]     \cr
  F_h(x)=  &\phi_\g(x)+\sum_{n=1}^\nu
                              \left[\phi_\g(x-x_n)+\phi_\g(x+x_n) \right] \cr
  N\rho_\delta(x)= &\sum_{j=1}^{M_r}\delta(x-\l_j) +\phi_\g(x)+
          \sum_{n=1}^\nu\left[\delta(x-x_h)+\delta(x+x_h)\right]\cr}  \eqn\defi
$$
and $K\ast$ is the convolution defined by
$$
       (K\ast f)(x)= \int_{-\infty}^{+\infty} dy
                       \, \phi_\g^\prime(x-y)f(y)                   \eqn\kconv
$$
The so--called ``density approach" consists in replacing, as $N\to\infty$,
both $\rho_N$ and $\rho_\delta$ with the same smooth function
$\rho$, representing the density of roots and holes on the real line.
This function is therefore the unique solution of the {\it linear} integral
equation (cft. eq.\deri)
$$
              2\pi\rho= F_\Th+N^{-1}(F_0+F_c-F_h)+ K\ast \rho      \eqn\linear
$$
which can be easily solved by Fourier transformation.
Combining this equation with eq.\deri, we now obtain, after some simple
manipulations
$$
        \rho_N=\rho+ G\ast\left(\rho_N-\rho_\delta\right)   \eqn\rewr
$$
where $G\ast=(2\pi+K)^{-1}\ast K\ast$ stands for the convolution with kernel
$$
        G(x)=\int{{dk}\o{2\pi}}e^{ikx}
           {{\sinh(\pi/2-\g)k}\o{\sinh(\pi-\g)k/2\,\cosh\g k/2}}  \eqn\gconv
$$
Finally, a simple application of the residue theorem to the analytic function
$\rho_N(1-e^{-iNZ_N})^{-1}$ plus an integration by
parts lead to the following formal {\it nonlinear} integral equation for the
counting function
$$
                     Z_N=Z + G\ast L_N                 \eqn\nonlin
$$
where $$
       L_N(x)=-iN^{-1}\log{{1-e^{iNZ_N(x+i0)}}\o{1-e^{-iNZ_N(x-i0)}}}
\eqn\lfun
$$
and $Z$ is the odd primitive of $2\pi\rho$, namely
$$
      Z=Z_\infty + N^{-1}(2\pi+K)\ast(F_0+F_c-F_h)                   \eqn\zzz
$$
with$$
     Z_\infty(x)=\left((2\pi+K)*F_\Th\right)(x)
         =2\arctan\left({\sinh\pi x/\g}\o{\cosh\pi\Th/\g}\right)   \eqn\zzinf
$$
Eq.\nonlin\ is a formal integral equation since the knowledge of
the exact position of holes and complex roots is required in $Z$. It becomes a
true integral equation for the ground state counting function. In any case
it is an exact expression satisfied by $Z_N$ where the number of site $N$
enters only in an explicit, parametric way, except for the positions
of the holes and of the
complex roots. After having fixed the corresponding quantum integers,
these parameters retain an implicit, mild dependence on $N$, for large $N$.

We shall now show that eq.\nonlin\ is very effective for establishing the
result \zapp, which is of crucial importance for the calculation of the
$S-$matrix. It is sufficent to check that the second nonlinear term
in the r.h.s. of eq.\nonlin\ is indeed of higher order in $N^{-1}$ relative
to the first. To this purpose observe that the integration contour of the
convolution in eq.\nonlin can be deformed away from the upper and lower edges
of the real axis, since  for sufficently large $N$ no complex roots can appear
in the whole strip $|{\rm Im\,}x|<\g/2$ and $G(x)$ is analytic there. Indeed
$Z_N$ tends to $Z_\infty$ as $N\to\infty$, and one can eplicitly check that the
imaginary part of $Z_\infty$ is positive definite for $0<{\rm Im\,}x<\g/2$ and
negative definite for $0>{\rm Im\,}x>-\g/2$. This also implies that the
contribution to the convolution integral is exponentially small in $N$ for all
values of the integration variable (let's call it $y$) where
${\rm Im\,}Z_\infty$ of order 1. For $|{\rm Re\,y}|$ of order log$\,N$,
we find ${\rm Im\,}Z_\infty(y)$ of order $N^{-1}$, so that the nonlinearity
$L_N$, rather than exponentially small, is also of order $N^{-1}$. But now the
exponential damping in $y$ of the kernel $G(x-y)$ guarantees that the
convolution integral is globally of order $N^{-2}$ or smaller, provided $|x|$
is kept smaller than $(\g/\pi)\log N$.  Hence we can write
$$
                         Z_N=Z+O(N^{-2})                      \eqn\approx
$$
Finally, the coefficent of the $N^{-1}$ term of $Z$, in eq.\zzz, can be
calculated in the $N\to\infty$ limit, with the techniques described at
length in paper I. After some straightforward albeit cumbersome algebra, this
yields$$
    \lim_{N\to\infty}\left((2\pi+K)\ast(F_0+F_c-F_h)\right)(x)=
           F(x\,;x_1,\ldots,\sub xM;\sub{\chi}1,\ldots,q\sub{\chi}M) \eqn\fff
$$
where the higher--level quantity $F$ is defined in eq.\shift. Together
with eqs. \approx\ and \zzz, this proves  eq.\zapp\ of section 3, as claimed.

Let us now consider the problem of calculating the energy of a given BA state,
through eq.\ene. We rewrite first the ``bare" energy function ${\rm e}_0(x)$
as$$
         {\rm e}_0(x)=-2\pi+\phi_{\g/2}(x+\Th)-\phi_{\g/2}(x-\Th)    \eqn\eee
$$
Then we calculate
$$\eqalign{
    \sum_{j=1} \phi_{\g/2}(x)=&\frac12 N\int_{-\infty}^{+\infty}
    \rho_\delta(x)\phi_{\g/2}(x)+\sum_{q=1}^{M_c}\phi_{\g/2}(u_q)
      -\sum_{h=1}^\nu\phi_{\g/2}(x_h)-\frac12\phi_{\g/2}(0) \cr
     =&\frac12 N\int_{-\infty}^{+\infty}\rho(x)\phi_{\g/2}(x+\Th)
         +\sum_{q=1}^{M_c}\phi_{\g/2}(u_q)
      -\sum_{h=1}^\nu\phi_{\g/2}(x_h)-\frac12\phi_{\g/2}(0)    \cr
     +&\frac12 N\int_{-\infty}^{+\infty}\left[\rho_\delta(x)-\rho_N(x)\right]
           \left((1-G)\ast\phi_{\g/2}\right)(x)    \cr}    \eqn\manip
$$
Through the residue theorem and an integration by parts, the last term can be
transformed, as done before for the counting function, into an integral of the
nonlinear term $L_N$, namely  the integral
$$
        \frac12 N\int_{-\infty}^{+\infty}{{L_N(x)}\o{\g\cosh\pi x/\g}}
$$
By the same argument used above, this last expression is globally of order
$N^{-1}$. Finally, inserting the explicit form of the continuum density
$\rho(x)$ into eq.\manip\ and recalling eqs. \ene\ and \eee, for the energy we
obtain eq.\enhole\ of the  main text
$$
      E=E_0+a^{-1}\sum_{h=1}^\nu {\rm e}(x_h) +O(a^{-1}N^{-1})       \eqn\etwo
$$
where
$$\eqalign{
    E_0=& -\pi a^{-1}N~+ \cr  &a^{-1}\int_0^\infty{{dk}\o k}
      {{\sinh(\pi-\g)k/2\,\sin k\Th}\o{\sinh\pi k/2\,\cosh\g k/2}}
     \left[2N\cos k\Th+1
             +{{\sinh(\pi-3\g)k/4}\o{\sinh(\pi-\g)k/4}}\right] \cr}
 \eqn\gse
$$
is the energy of the ground state.

We would like to close this appendix with a comment on the limitations of the
density approach, where one deals only with the solution $\rho$ of the linear
equation \linear. Regarding $\rho(x)$ as the actual density of real roots and
holes in the $N\to\infty$ limit, it is natural to use it to replace summations
with integrals. What one learns from the exact treatment presented above as
well as from computer calculations, is that the error made in such a
replacement depends crucially on the large $x$ behaviour of the quantity which
is to be summed. This error is down by $N^{-1}$ only when there is exponential
damping in $x$. This means, for instance, that the integral of $\rho(x)$ does
not reproduce in general the exact number of real roots and holes, but rather
some $\g-$dependent quantity close to it. Misundertanding this for the actual
number of real roots and holes would lead to the absurd result that the holes
have a $\g-$dependent value of the $SU(2)_q$ spin, which is instead necessarily
integer or half--integer and, in the particular case of the holes, just 1/2 for
any value of $\g$.

\refout
\endpage
\chapter{Figure Captions}

\item{Fig.1.}
Graphical representation of the $R$-matrix.
\bigskip
\item{Fig.2.}
The standard row-to-row monodromy matrix. The numbers from 1 to N
label the vertical spaces and each vertical line represents a couple of
free indices
in the corresponding vertical space. Indices on the internal
horizontal lines are summed over.
\bigskip
\item{Fig.3.}
The doubled monodromy matrix used to describe systems with fixed
boundary conditions. The correct contraction over the internal indices
is dictated by the position of the arcs.
\bigskip
\item{Fig.4.}
Graphical representation of the compatibility relations \matk\ between the
two-body scattering and reflections on the left wall.
\bigskip
\item{Fig.5.}
Graphical representation of the compatibility relations \matkp\ between the
two-body scattering and reflections on the right wall.
\bigskip
\item{Fig.6.}
The unit time evolution operator $U(\Th)$ for even N (top) and odd N
(bottom). As usual, indices over internal lines are summed over.
\bigskip
\item{Fig.7.}
Reconstruction of the diagonal light-cone lattice through powers of
the unit time evolution operator (case N even).
\bigskip
\item{Fig.8.}
Assignement of the local height variables on the plaquettes
cut by the time-zero and time-one lines. As is evident from the figure,
these two lines sandwich the unit time evolution operator  $U(\Th)$.
\bigskip
\item{Fig.9.}
The Bratteli diagram corresponding to the case N=6. Right-moving
paths on this diagram arriving at height $j$ define the basis in the
space of irreps of weight $j$.
\bigskip
\item{Fig.10.}
The ``ground state dominating" configuration of local height
variables.
\bigskip
\item{Fig.11.}
Graphical representation of the vertex--face correspondence. The strip
sandwiched between the two time lines represents the unit time
evolution.
\bigskip
\item{Fig.12.}
Graphical representation of the one--soliton operator corresponding to the
particle with rapidity $\t_2$ in a system with $\nu=4$ particles. At each
intersection, the appropriate two--body $S-$matrix acts. In the collisions
against the wall, $\t_2$ is flipped and the two--by--two matrix $g_2(\pm\t_2)$
acts on the internal states of the soliton. The choice
of $\t_2$ as largest rapidity is done purely for graphical
convenience.
\bigskip
\item{Fig.13.}
One of the local height configurations that dominate the one--hole BA state.
of spin $J=1$.
\bigskip
\item{Fig.14.}
The ``renormalized" version of Fig.13.
\bigskip
\item{Fig.15.}
Flipping the last portion of the configuration from the solid  to the dotted
line does not change the quantum spin $J=p/2-1$, but tranform the type II state
into type I.
\bigskip
\item{Fig.16.}
Plot of the $N/2\pi$ times the counting function $Z_N(x)$ versus $\tanh x$ for
$N=64$, $J=1$, $\Th=.15$ and various values of $\g$ from $\pi/6$ to
$0.999\pi/4$.  The quantum integers associates to the two holes are ${\bar
I}_1=8$ and ${\bar I}_1=21$. The critical value of $(N/2\pi)Z_N$ is $N/2+J=33$,
while that of $\g$ is, roughly, $\g^\ast=0.21\pi$.
\bigskip

\bye